\documentclass[aps,groupedaddress,showpacs,floatfix]{revtex4}
\usepackage{amssymb}
\usepackage{amsmath}
\usepackage{graphicx}
\DeclareMathOperator{\sgn}{sgn}
\DeclareMathOperator{\Ai}{Ai}
\begin{document}

\title{Two-time free energy distribution function in (1+1) directed polymers}

\author{Victor Dotsenko}

\affiliation{LPTMC, Universit\'e Paris VI, 75252 Paris, France}

\affiliation{L.D.\ Landau Institute for Theoretical Physics,
 119334 Moscow, Russia}

\date{\today}

\begin{abstract}
Two time free energy
distribution function in one-dimensional directed polymers
in random potential is derived  in terms of the Bethe ansatz replica
technique by mapping the replicated problem to the $N$-particle quantum boson
system with attractive interactions.

\end{abstract}


\maketitle

\medskip

\section{Introduction}

We consider the model of directed polymers defined in terms of an elastic string $\phi(\tau)$
directed along the $\tau$-axes within an interval $[0,t]$ which passes through a random medium
described by a random potential $V(\phi,\tau)$. The energy of a given polymer's trajectory
$\phi(\tau)$ is
\begin{equation}
   \label{1}
   H[\phi(\tau), V] = \int_{0}^{t} d\tau
   \Bigl\{\frac{1}{2} \bigl[\partial_\tau \phi(\tau)\bigr]^2
   + V[\phi(\tau),\tau]\Bigr\};
\end{equation}
where the disorder potential $V[\phi,\tau]$
is Gaussian distributed with a zero mean $\overline{V(\phi,\tau)}=0$
and the $\delta$-correlations
${\overline{V(\phi,\tau)V(\phi',\tau')}} = u \delta(\tau-\tau') \delta(\phi-\phi')$
The parameter $u$ describes the strength of the disorder.

The system of such type as well as the equivalent problem of the KPZ-equation
\cite{KPZ} describing the growth in time of an interface
in the presence of noise
have been the subject of intense investigations during the past almost three
decades (see e.g.
\cite{hh_zhang_95,burgers_74,kardar_book,hhf_85,numer1,numer2,kardar_87,bouchaud-orland,Brunet-Derrida,
Johansson,Prahofer-Spohn,Ferrari-Spohn1}).
The breakthrough in these studies took place in 2010 when the exact solution for the free energy
probability distribution function (PDF) for the model with fixed boundary condition has been found
\cite{KPZ-TW1a,KPZ-TW1b,KPZ-TW1c,KPZ-TW2,BA-TW1,BA-TW2,BA-TW3,LeDoussal1}.
It was show that this PDF is given by the Tracy-Widom (TW) distribution of the largest eigenvalue of
the Gaussian Unitary Ensamble (GUE) \cite{TW-GUE}.
Since that time important progress in understanding of the statistical properties of the KPZ-class
systems has been achieved (for the review see \cite{Corwin,Borodin}).
In particular, by this time it is shown that the free energy PDF of the directed polymer model
(\ref{1}) with free boundary conditions is given by the Gaussian Orthogonal Ensemble (GOE) TW distribution
\cite{LeDoussal2,goe}, while in the presence of a "wall" ($\phi(\tau) \geq 0, \; 0 \leq \tau \leq t$)
such PDF is given by the Gaussian Simplectic Ensemble (GSE) TW distribution \cite{LeDoussal3}.
Besides, the two-point free energy distribution function
which describes joint statistics of the free energies of the directed polymers
coming to two different endpoints has been derived in  \cite{Prolhac-Spohn}
and quite recently the explicit expression for the PDF for the end-point $\phi(t)$ fluctuations
has been obtained \cite{math1, math2, math3, end-point}.

All these studies, however, describe the statistics of the model (\ref{1}) in the so called
"one-time" situation. In this paper I am going to derive the joint probability distribution function,
$W(f_{1}, f_{2}, \Delta)$,
for the free energies $f_{1}$ and $f_{2}$ of two directed polymers with fixed boundary conditions
at two different times, $\phi(t_{1}) = 0$ and $\phi(t_{2}) = 0$, with
$t_{1} = t$, $t_{2} = (1+\Delta)t$ in the limit $t\to\infty$ when the parameter $\Delta > 0$
remains finite. The derivation is done in terms of the Bethe ansatz replica technique.
The main points of this approach are described in Section II
(for the details of the method see e.g. \cite{BA-TW3,goe,end-point}).
Detailed derivation of $W(f_{1}, f_{2}, \Delta)$ is given in Section III.
Unfortunately, at present stage it is expressed in terms rather
complicated "determinant-like" object, eq.(\ref{76}),
which is {\it not} the Fredholm determinant, and whose analytic properties are still to be studied,
although in the limit cases it reduces to the predictable results, namely:
 (1) in the limit $f_{1}\to -\infty$ one recovers
the GUE TW distribution for $f_{2}$;
(2) in the limit $f_{2}\to -\infty$ one recovers
the GUE TW distribution for $f_{1}$;
and (3) in the limit $\Delta\to \infty$ one obtains
two independent GUE TW distribution for $f_{1}$ and $f_{2}$.

\section{Bethe ansatz replica technique}

For the fixed boundary conditions, $\phi(0) = \phi(t) = x$, the partition function
of the model (\ref{1}) is
\begin{equation}
\label{2}
   Z_{t}(x) = \int_{\phi(0)=0}^{\phi(t)=x}
              {\cal D} \phi(\tau)  \;  \mbox{\Large e}^{-\beta H[\phi]}
\; = \; \exp\bigl(-\beta F_{t}(x)\bigr)
\end{equation}
where $\beta$ is the inverse temperature and $F_{t}(x)$ is the free energy.
In the limit $t\to\infty$ the free energy scales as
$\beta F_{t}(x) = \beta f_{0} t + \beta x^{2}/2t + \lambda_{t} f(x)$,
where $f_{0}$ is the selfaveraging free energy density,
$\lambda_{t} \propto t^{1/3}$ and $f(x)$ is a random quantity.
It is the statistics of $f(x)$ which in the limit $t\to\infty$ is
expected to be described by a non-trivial universal distribution $W(f)$.
In fact the first two trivial terms of this free energy can be easily
eliminated by simple redefinition of the partition function,
and therefore, to simplify the formulas, these two terms in what follows will be
just omitted.

The calculation of the probability distribution function
\begin{equation}
\label{3}
W(f) \; = \; \lim_{t\to\infty} \;
\mbox{Prob}\bigl[f(x) \; > \; f\bigr]
\end{equation}
is performed in terms of the generating function
\begin{equation}
\label{4}
W(f) = \lim_{t\to\infty}
\sum_{N=0}^{\infty} \frac{(-1)^{N}}{N!}
\exp\bigl(\lambda_{t} N f\bigr) \;
\overline{\bigl[Z_{t}(x)\bigr]^{N}}
\end{equation}
where $\overline{(...)}$ denotes the averaging over the random potential $V(\phi,\tau)$.
Instead of the one-point replica partition function $\overline{\bigl[Z_{t}(x)\bigr]^{N}}$
 one can introduce more general $N$-point object:
\begin{equation}
\label{5}
\Psi(x_{1}, ..., x_{N} ; t) \; \equiv \;
\overline{Z_{t}(x_{1}) \, Z_{t}(x_{2}) \, ... \, Z_{t}(x_{N})} =
\prod_{a=1}^{N} \Biggl[\int_{\phi_a(0)=0}^{\phi_a(t)=x_a} {\cal D} \phi_a(\tau)\Biggr]
  \;  \overline{\Biggl(\exp\bigl[-\beta \sum_{a=1}^{N} H [\phi_{a}] \bigr]\Biggr)}
\end{equation}
It can be easily shown that $\Psi({\bf x}; t)$ is the wave function of $N$-particle
boson system with attractive $\delta$-interaction:
\begin{equation}
   \label{6}
\beta \, \partial_t \Psi({\bf x}; t) =
\frac{1}{2}\sum_{a=1}^{N}\partial_{x_a}^2 \Psi({\bf x}; t)
+\frac{1}{2}\, \kappa \sum_{a\not=b}^{N} \delta(x_a-x_b) \; \Psi({\bf x}; t)
\end{equation}
(where $\kappa = \beta^{3} u$) with the initial condition $\Psi({\bf x}; 0) = \Pi_{a=1}^{N} \delta(x_a)$.
The wave function  $\Psi({\bf x}; t)$ of this quantum problem can be represented in terms of the linear combination
of the corresponding eigenfunctions  of eq.(\ref{6}).
A generic  eigenstate of such system is characterized by $N$ momenta
$\{ Q_{a} \} \; (a=1,...,N)$ which split into
$M$  ($1 \leq M \leq N$) "clusters" each described by
continuous real momenta $q_{\alpha}$ $(\alpha = 1,...,M)$
and characterized by $n_{\alpha}$ discrete imaginary "components"
(for details see \cite{Lieb-Liniger,McGuire,Yang,Calabrese,rev-TW}):
\begin{equation}
   \label{8}
Q_{a} \; \to \; q^{\alpha}_{r} \; = \;
q_{\alpha} - \frac{i\kappa}{2}  (n_{\alpha} + 1 - 2r)
\;\; ; \; \;\;\; \;\;\; \;\;\;
(r = 1, ..., n_{\alpha}\,; \; \; \alpha = 1, ..., M)
\end{equation}
with the global constraint $\sum_{\alpha=1}^{M} n_{\alpha} = N$.
Explicitly,
\begin{equation}
\label{9}
\Psi_{{\bf Q}}({\bf x}) =
\sum_{{\cal P}}  \;
\prod_{1\leq a<b}^{N}
\Biggl[
1 +i \kappa \frac{\sgn(x_{a}-x_{b})}{Q_{{\cal P}_a} - Q_{{\cal P}_b}}
\Biggr] \;
\exp\Bigl[i \sum_{a=1}^{N} Q_{{\cal P}_{a}} x_{a} \Bigr]
\end{equation}
where the vector ${\bf Q}$ denotes the set of all $N$ momenta eq.(\ref{8}) and
the summation goes over $N!$ permutations ${\cal P}$ of $N$ momenta $Q_{a}$,
 over $N$ particles $x_{a}$.
In terms of the above eigenfunctions the solution of eq.(\ref{6}) can be expressed as follows:
\begin{equation}
\label{7}
\Psi({\bf x}; t) = \frac{1}{N!} \int {\cal D}_Q \; |C({\bf Q})|^{2} \; \Psi_{\bf Q}({\bf x}) \Psi^{*}_{\bf Q}(0) \;
\exp\bigl(-t E({\bf Q}) \bigr)
\end{equation}
where
the symbol $\int {\cal D}_Q$ denotes the integration over $M$ continuous parameters $\{ q_{1}, ..., q_{M}\}$,
the summations over $M$ integer parameters $\{ n_{1}, ..., n_{M}\}$ as well as summation over $M=1,..,N$.
$|C({\bf Q})|^{2}$ is the normalization factor,
\begin{eqnarray}
   \nonumber
|C({\bf Q})|^{2} &=& \frac{\kappa^{N}}{\prod_{\alpha=1}^{M}\bigl(\kappa n_{\alpha}\bigr)}
\prod_{\alpha<\beta}^{M}
\frac{\big|q_{\alpha}-q_{\beta} -\frac{i\kappa}{2}(n_{\alpha}-n_{\beta})\big|^{2}}{
      \big|q_{\alpha}-q_{\beta} -\frac{i\kappa}{2}(n_{\alpha}+n_{\beta})\big|^{2}}
\\
\nonumber
\\
&=&
\kappa^{N} \det\Biggl[
   \frac{1}{\frac{1}{2}\kappa n_{\alpha} - i q_{\alpha}
          + \frac{1}{2}\kappa n_{\beta} + iq_{\beta}}\Biggr]_{\alpha,\beta=1,...M}
 \label{10}
\end{eqnarray}
and $E({\bf Q})$ is the eigenvalue (energy) of the
eigenstate $\Psi_{\bf Q}({\bf x})$,
\begin{equation}
\label{11}
E({\bf Q}) \; = \;
\frac{1}{2\beta} \sum_{\alpha=1}^{N} Q_{a}^{2} =
 \frac{1}{2\beta} \sum_{\alpha=1}^{M} \; n_{\alpha} q_{\alpha}^{2}
- \frac{\kappa^{2}}{24\beta}\sum_{\alpha=1}^{M} n_{\alpha}^{3}
\end{equation}
In this way for the distribution function (\ref{4}) one gets the following expression
\begin{equation}
\label{12}
W(f) = \lim_{t\to\infty}\Biggl\{
1 + \sum_{N=1}^{\infty} \frac{(-1)^{N}}{N!}
\exp\bigl(\lambda_{t} N f\bigr) \;
\Psi({\bf x}; t)\Big|_{x_{a} = x} \Biggr\}
\end{equation}
where the wave function $\Psi({\bf x}; t)$ is defined by eqs.(\ref{7})-(\ref{11}). Performing the summations
over integer parameters $\{n_{1}, ..., n_{M}\}$ , integrating over  continuous parameters $\{q_{1}, ..., q_{M}\}$
and summing over $M$ and $N$,
in the limit $t\to\infty$ one eventually obtains the GUE Tracy-Widom distribution
in the form of the Fredholm determinant \cite{BA-TW2,BA-TW3,LeDoussal1}:
\begin{equation}
 \label{13}
W(f) = F_{2}\bigl(- f/2^{2/3}\bigr) \equiv \det\bigl[\hat{1} - \hat{K}_{-f/2^{2/3}}\bigr]
\end{equation}
where $\hat{K}_{s}$ is the integral operator on $[0, +\infty)$ with the Airy kernel:
\begin{equation}
 \label{14}
K_{s}(\omega, \omega') = \int_{0}^{+\infty} dy \Ai\bigl(y +\omega + s\bigr) \Ai\bigl(y +\omega' + s\bigr)
\; \; \; \; \; \; (\omega, \omega' \geq 0)
\end{equation}

\section{Two-time probability distribution function}

\subsection{Replica Bethe ansatz definition}

Let us consider the situation when one polymer trajectory is arriving to zero at time $t_{1} = t$ and
having (random) free energy $f_{t_{1}}(0)$,
while the other trajectory is arriving to zero at time $t_{2} = (1+\Delta) t$ and
having (random) free energy $f_{t_{2}}(0)$. One would like to compute the joint probability distribution function
\begin{equation}
\label{15}
W(f_{1}, f_{2}, \Delta) =
\lim_{t\to\infty} \;
\mbox{Prob}\bigl[f_{t_{1}}(0) > f_{1}; \; f_{t_{2}}(0) > f_{2} \bigr]
\end{equation}
In terms of the replica partition functions this quantity can be defined as follows:
\begin{equation}
\label{16}
W(f_{1}, f_{2}, \Delta) = \lim_{t\to\infty}
\sum_{K=0}^{\infty} \sum_{N=0}^{\infty}
\frac{(-1)^{K+N}}{K! \; N!}
\exp\bigl(\lambda_{1} K f_{1} + \lambda_{2} N f_{2}\bigr) \;
\overline{Z_{t_{1}}^{K}(0) \, Z_{t_{2}}^{N}(0)}
\end{equation}
where
\begin{eqnarray}
 \label{17}
\lambda_{1} &=& \frac{1}{2} \bigl(\beta^{5} u^{2} t\bigr)^{1/3} \; \equiv \; \lambda_{t}
\\
\nonumber
\\
\label{18}
\lambda_{2} &=& \frac{1}{2} \bigl(\beta^{5} u^{2} (1+\Delta) t\bigr)^{1/3} \; = \; (1+\Delta)^{1/3} \lambda_{t}
\end{eqnarray}

\begin{eqnarray}
\label{19}
   Z_{t_{1}}(0) &=& \int_{\phi(0)=0}^{\phi(t_{1})=0}
              {\cal D} \phi(\tau)  \;  \mbox{\Large e}^{-\beta H[\phi]}
\\
\nonumber
\\
\label{20}
   Z_{t_{2}}(0) &=& \int_{\phi(0)=0}^{\phi(t_{2})=0}
              {\cal D} \phi(\tau)  \;  \mbox{\Large e}^{-\beta H[\phi]}
\; = \;
\int_{-\infty}^{+\infty} dx \;
Z_{t_{1}}(x)\, Z^{*}_{(t_{2}-t_{1})}(x)
\end{eqnarray}
and
\begin{equation}
 \label{21}
Z^{*}_{(t_{2}-t_{1})}(x) \; = \;
\int_{\phi(t_{1})=x}^{\phi(t_{2})=0}
              {\cal D} \phi(\tau)  \;  \mbox{\Large e}^{-\beta H[\phi]}
\end{equation}
is the partition function of the directed polymer system in which time goes backwards,
from $t_{2}$ to $t_{1}$.
For technical reasons (for proper regularization of the integration over $x$ at
$\pm$ infinities)
it is convenient to split the partition function
$Z_{t_{2}}$ into two parts, the "left" and the "right" ones:
\begin{equation}
 \label{22}
Z_{t_{2}}(0) \; = \;
\int_{-\infty}^{0} dx \;
Z_{t_{1}}(x)\, Z^{*}_{(t_{2}-t_{1})}(x)
\; + \;
\int_{0}^{+\infty} dx \;
Z_{t_{1}}(x)\, Z^{*}_{(t_{2}-t_{1})}(x)
\end{equation}
Substituting eqs.(\ref{17})-(\ref{22}) (with $t_{1} = t$ and $(t_{2}-t_{1}) = \Delta \, t$)
into eq.(\ref{16}) and taking into account the definition (\ref{5})
we get:
\begin{eqnarray}
\label{23}
W(f_{1}, f_{2}, \Delta) &=& \lim_{t\to\infty} \Biggl\{
\sum_{L,K,R=0}^{\infty}
\frac{(-1)^{L+K+R}}{L!\, K!\, R!} \,
\exp\Bigl[\lambda_{t} K f_{1} + (1+\Delta)^{1/3} \lambda_{t} (L+R) f_{2}\Bigr]
\times
\\
\nonumber
\\
\nonumber
&\times&
\int_{-\infty}^{0} dx_{1}...dx_{L} \int_{0}^{+\infty} dy_{R} ... dy_{1}
\Psi\bigl(x_{1},...,x_{L}, \underbrace{0, ..., 0}_{K}, y_{R}, ..., y_{1} ; \, t\bigr) \;
\Psi^{*}\bigl(x_{1},...,x_{L}, y_{R}, ..., y_{1} ; \, \Delta \, t\bigr)
\Biggr\}
\end{eqnarray}
where the second (conjugate) wavefunction represent the "backward" propagation
from the time moment $t_{2} = (1+\Delta) t$ to the previous time moment $t_{1}= t$.
Schematically the above expression is represented in Figure 1.

\begin{figure}[h]
\begin{center}
   \includegraphics[width=12.0cm]{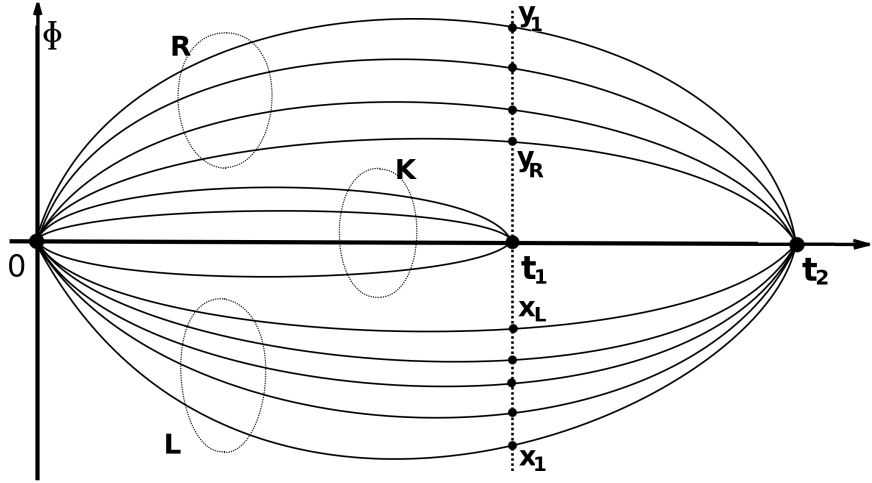}
\caption[]{Schematical representation of the directed polymer paths
corresponding to eq.(\ref{23})}
\end{center}
\label{figure1}
\end{figure}

\subsection{Calculations}

Substituting the representations (\ref{7}) and (\ref{9}) for the wave functions
$\Psi\bigl({\bf x}, {\bf 0}, {\bf y} ; \, t\bigr)$ and
$\Psi^{*}\bigl({\bf x},  {\bf y} ; \, \Delta \, t\bigr)$ in eq.(\ref{23}) we obtain:
\begin{eqnarray}
\nonumber
W(f_{1}, f_{2}, \Delta) &=& \lim_{t\to\infty} \Biggl\{
\sum_{L,K,R=0}^{\infty}
\frac{(-1)^{L+K+R}}{L!\, K!\, R!} \,
\exp\Bigl[\lambda_{t} K f_{1} + (1+\Delta)^{1/3} \lambda_{t} (L+R) f_{2}\Bigr]
\times
\\
\nonumber
\\
&\times&
\int{\cal D}_Q \; \int {\cal D}_P \;
|C({\bf Q})|^{2} |C({\bf P})|^{2}\; \exp\bigl[-t E({\bf Q})-\Delta t E({\bf P}) \bigr]
\; I({\bf Q}, {\bf P})
\Biggr\}
\label{24}
\end{eqnarray}
where
\begin{eqnarray}
\nonumber
I\bigl({\bf Q}; {\bf P} \bigr) &=&
\sum_{{\cal P}}  \;\sum_{\tilde{{\cal P}}} \;
\Pi_{LK}({\bf Q}) \; \Pi_{LR}({\bf Q}) \; \Pi_{KR} ({\bf Q}) \; \Pi^{*}_{LR} ({\bf P})
\times
\\
\nonumber
\\
\nonumber
&\times&
\int_{-\infty}^{0} dx_{1}...dx_{L}
\prod_{1\leq a<b}^{L}
\Biggl[
\Bigl(
1 + i \kappa \frac{\sgn(x_{a}-x_{b})}{Q_{{\cal P}_a} - Q_{{\cal P}_b}}
\Bigr)
\Bigl(
1 -i \kappa \frac{\sgn(x_{a}-x_{b})}{P_{\tilde{{\cal P}}_a} - P_{\tilde{{\cal P}}_b}}
\Bigr)
\Biggr]
\exp\Bigl[i \sum_{a=1}^{L} \bigl(O_{{\cal P}_{a}} - P_{\tilde{{\cal P}}_{a}} -i\epsilon\bigr) x_{a} \Bigr]
\times
\\
\nonumber
\\
\nonumber
&\times&
\int_{0}^{+\infty} dy_{R}...dy_{1}
\prod_{1\leq c<d}^{R}
\Biggl[
\Bigl(
1 - i \kappa \frac{\sgn(y_{c}-y_{d})}{Q_{{\cal P}_{(L+K+c)}} - Q_{{\cal P}_{(L+K+d)}}}
\Bigr)
\Bigl(
1 + i \kappa \frac{\sgn(y_{c}-y_{d})}{P_{\tilde{{\cal P}}_{(L+c)}} - P_{\tilde{{\cal P}}_{(L+d)}}}
\Bigr)
\Biggr]
\times
\\
\nonumber
\\
&\times&
\exp\Bigl[i \sum_{c=1}^{R} \bigl(Q_{{\cal P}_{(L+K+c)}} - P_{\tilde{{\cal P}}_{(L+c)}} + i\epsilon\bigr) y_{c} \Bigr]
\label{25}
\end{eqnarray}
Here, to regularize the integrations at $\pm$ infinities the supplementary factors $\pm i\epsilon$
are introduced (which will be set to zero in the final result). The factors
$\Pi_{ij}({\bf Q}) \; (i,j = LK, LR, KR)$ and $\Pi^{*}_{LR} ({\bf P})$ in the above equation are the
cross-products of the prefactors of the type $[1-i\kappa/(Q_{(i)}-Q_{(j)})]$ over the particles
belonging to the sectors "L" and "K", "L" and "R" and "K" and "R" (note that $P_{KK}({\bf Q}) \equiv 1$).
Correspondingly, $\Pi^{*}_{LR} ({\bf P})$ is the cross-product of the prefactors of the type
$[1+i\kappa/(P_{(i)}-P_{(j)})]$ over the particles
belonging to the sectors "L" and "R".

The summation over all permutations ${\cal P}$ of $(L+K+R)$ momenta $Q_{a}$ over $(L+K+R)$ particles
of the sectors "L" $(x_{1}, ..., x_{L})$, "K" (all coordinates equal to zero) and "R" $(y_{R}, ..., y_{1})$
can be split into four parts: \\
(1) the summation over permutations ${\cal P}^{(L)}$ of $L$ (taken at random) momenta $q$ over
$L$ particles of the sector "L";\\
(2) the summation over permutations ${\cal P}^{(R)}$ of $R$ (taken at random) momenta $q$ over
$R$ particles of the sector "R";\\
(3) the summation over permutations ${\cal P}^{(K)}$ of $K$ remaining momenta $q$ over
$K$ particles of the central sector "K";\\
(4) the summation over permutations ${\cal P}^{(LKR)}$ of the momenta $q$ over sectors
"L", "K" and "R".

Similar splitting can be done for the summation over the permutations $\tilde{{\cal P}}$ of
$(L+R)$ momenta $P_{a}$ over $(L+R)$ particles of the sectors "L" and "R".
In other words, the summations over the permutations in eq.(\ref{25}) can be represented as follows:
\begin{equation}
 \label{26}
\sum_{{\cal P}}  \;\sum_{\tilde{{\cal P}}} \; \Bigl( ... \Bigr) \; = \;
\sum_{{\cal P}^{(LKR)}} \sum_{\tilde{{\cal P}}^{(LR)}} \;
\sum_{{\cal P}^{(L)}} \; \sum_{\tilde{{\cal P}}^{(L)}} \;
\sum_{{\cal P}^{(R)}} \; \sum_{\tilde{{\cal P}}^{(R)}} \;
\sum_{{\cal P}^{(K)}} \;
\Bigl( ... \Bigr)
\end{equation}

To perform the summations over all these permutations we can use the following important property of the Bethe
ansatz wave function, eq.(\ref{9}). Namely, it has such structure that for ordered particle's positions
in the summation over the permutations the momenta components $q^{\alpha}_{r}$ belonging to the
same cluster also remain ordered. In other words, if we consider the momenta of a cluster $\alpha$,
$\{q^{\alpha}_{1}, ... , q^{\alpha}_{n_{\alpha}}\}$, eq.(\ref{8}), belonging correspondingly to the particles
$\{x_{i_{1}} < x_{i_{2}} < ... < x_{i_{n_{\alpha}}}\}$, the permutation of any two momenta
$q^{\alpha}_{r}$ and $q^{\alpha}_{r'}$ of this {\it ordered } set gives zero contribution.
In our case we have three groups of particles:
$\{x_{1}, ..., x_{L}\} \; < \; \{0, ..., 0\} \; < \;  \{y_{R}, ..., y_{1}\}$.
Thus, in order to perform the summation over the permutations ${\cal P}^{(LKR)}$
it is sufficient to split the momenta of each cluster into three parts:
\begin{equation}
 \label{27}
\{q^{\alpha}_{1}, ... , q^{\alpha}_{n_{\alpha}}\} \; \to \;
\{ q^{\alpha}_{1}, ..., q^{\alpha}_{m_{\alpha}} \, \big|\big| \,
    q^{\alpha}_{m_{\alpha}+1}, ..., q^{\alpha}_{m_{\alpha}+k_{\alpha}} \, \big|\big| \,
    q^{\alpha}_{m_{\alpha}+k_{\alpha}+1}, ..., q^{\alpha}_{m_{\alpha}+k_{\alpha}+s_{\alpha}} \}
\end{equation}
defined by three integer parameters $m_{\alpha}, k_{\alpha}$ and $s_{\alpha}$, such that
$m_{\alpha} + k_{\alpha} + s_{\alpha} = n_{\alpha}$. In such cluster structure
the momenta components of the left group
$\{q^{\alpha}_{1}, ..., q^{\alpha}_{m_{\alpha}}\}$
go to the particles of the sector "L";
the momenta components of the central group
$\{q^{\alpha}_{m_{\alpha}+1}, ..., q^{\alpha}_{m_{\alpha}+k_{\alpha}}\}$
go to the particles of the sector "K";
and
the momenta components of the right group
$\{q^{\alpha}_{m_{\alpha}+k_{\alpha}+1}, ..., q^{\alpha}_{m_{\alpha}+k_{\alpha}+s_{\alpha}}\}$
go to the particles of the sector "R".

Similarly, to perform the summation over the permutations $\tilde{{\cal P}}^{(LR)}$
of the momenta $p^{\alpha}_{r}$ each cluster is split into two parts:
$\{ p^{\alpha}_{1}, ..., p^{\alpha}_{m_{\alpha}} \, \big|\big| \,
    p^{\alpha}_{m_{\alpha}+1}, ..., p^{\alpha}_{m_{\alpha}+s_{\alpha}} \}$.
Here the components
$\{p^{\alpha}_{1}, ..., p^{\alpha}_{m_{\alpha}}\}$
go to the particle of the sector "L", while the components
$\{ p^{\alpha}_{m_{\alpha}+1}, ..., p^{\alpha}_{m_{\alpha}+s_{\alpha}} \}$
go to the particle of the sector "R".
In this way, the summations over the permutations ${\cal P}^{(LKR)}$ and $\tilde{{\cal P}}^{(LR)}$
is changed by the summations over the integer parameters $\{m_{\alpha}, k_{\alpha},s_{\alpha}\}$
constrained by the conditions:
\begin{equation}
\label{28}
\sum_{\alpha=1}^{M} \; m_{\alpha}  =  L \, ,
\; \; \; \; \; \; \;
\sum_{\alpha=1}^{M} \; k_{\alpha} =   K \, ,
\; \; \; \; \; \; \;
\sum_{\alpha=1}^{M} \; s_{\alpha}  =  R
\end{equation}

Performing simple integrations over $x_{1}, ..., x_{L}$ and $y_{R}, ..., y_{1}$  in eq.(\ref{25})
and taking into account that all cross-product factors $\Pi_{ij}$ are symmetric with respect to the
permutations ${\cal P}^{(L)}, {\cal P}^{(K)}, {\cal P}^{(R)}, \tilde{{\cal P}}^{(L)}$ and 
$\tilde{{\cal P}}^{(R)}$,
we get:
\begin{equation}
 \label{29}
I({\bf Q}, {\bf P}) = L! \, K! \, R!
\prod_{\alpha=1}^{M}\Biggl[\sum_{m_{\alpha}+k_{\alpha}+s_{\alpha}\geq 1}\Biggr]
\delta_{\sum m_{\alpha}, L} \; \delta_{\sum k_{\alpha}, K} \; \delta_{\sum s_{\alpha}, R} \;
{\cal G}\bigl({\bf Q}, {\bf P}\bigr)  \;
D_{L}({\bf Q},{\bf P}) \, D_{R}({\bf Q},{\bf P})
\end{equation}
where $\delta_{i,j}$ is the kronecker symbol,
\begin{equation}
 \label{38}
{\cal G}\bigl({\bf Q}, {\bf P}\bigr)   =
\Pi_{LK} \bigl({\bf Q}\bigr) \; \Pi_{LR} \bigl({\bf Q}\bigr)\; \Pi_{KR}\bigl({\bf Q}\bigr)  \; \Pi^{*}_{LR}\bigl({\bf P}\bigr)
\end{equation}
and
\begin{equation}
 \label{30}
D_{L}({\bf Q},{\bf P}) =
\sum_{{\cal P}^{(L)}}  \sum_{\tilde{{\cal P}}^{(L)}}
\frac{1}{
\prod_{a=1}^{L}
\Bigl[\sum_{b=1}^{a}
\bigl(Q_{{\cal P}^{(L)}_{b}} - P_{\tilde{{\cal P}}^{(L)}_{b}} -i\epsilon\bigr)
\Bigr]}
\prod_{1\leq a<b}^{L}
\Biggl[\Bigl(
1-\frac{i\kappa}{Q_{{\cal P}^{(L)}_{a}} - Q_{{\cal P}^{(L)}_{b}}}
\Bigr)
\Bigl(
1+\frac{i\kappa}{P_{\tilde{{\cal P}}^{(L)}_{a}} - P_{\tilde{{\cal P}}^{(L)}_{b}}}
\Bigr)
\Biggr]
\end{equation}
\begin{eqnarray}
 \nonumber
D_{R}({\bf Q},{\bf P}) &=&
\sum_{{\cal P}^{(R)}}  \sum_{\tilde{{\cal P}}^{(R)}}
\frac{1}{
\prod_{a=L+K+R}^{L+K+1}
\Bigl[\sum_{b=L+K+R}^{a}
\bigl(Q^{*}_{{\cal P}^{(L)}_{b}} - P^{*}_{\tilde{{\cal P}}^{(L)}_{b}} +i\epsilon\bigr)
\Bigr]}
\times
\\
\nonumber
\\
&\times&
\prod_{L+K+1\leq a<b}^{L+K+R}
\Biggl[\Bigl(
1+\frac{i\kappa}{Q^{*}_{{\cal P}^{(L)}_{a}} - Q^{*}_{{\cal P}^{(L)}_{b}}}
\Bigr)
\Bigl(
1-\frac{i\kappa}{P^{*}_{\tilde{{\cal P}}^{(L)}_{a}} - P^{*}_{\tilde{{\cal P}}^{(L)}_{b}}}
\Bigr)
\Biggr]
\label{31}
\end{eqnarray}
Here the momenta components in the "L" sector are
\begin{eqnarray}
\nonumber
Q_{a} &\equiv& q^{\alpha}_{r} = q_{\alpha} -\frac{1}{2} i \kappa \bigl( m_{\alpha}+k_{\alpha}+s_{\alpha} +1 - 2r\bigr)
\\
\label{32}
\\
\nonumber
P_{a} &\equiv& p^{\alpha}_{r} = p_{\alpha} +\frac{1}{2} i \kappa \bigl( m_{\alpha}+s_{\alpha} +1 - 2r\bigr)
\end{eqnarray}
with $r = 1, ..., m_{\alpha}$, while in the "R" sector  the momenta components
are counted "backwards" which makes them complex conjugate:
\begin{eqnarray}
\nonumber
Q^{*}_{a} &\equiv& {q^{*}}^{\alpha}_{r} = q_{\alpha} + \frac{1}{2} i \kappa \bigl( m_{\alpha}+k_{\alpha}+s_{\alpha} +1 - 2r\bigr)
\\
\label{33}
\\
\nonumber
P^{*}_{a} &\equiv& {p^{*}}^{\alpha}_{r} = p_{\alpha} -\frac{1}{2} i \kappa \bigl( m_{\alpha}+s_{\alpha} +1 - 2r\bigr)
\end{eqnarray}
with $r = 1, ..., s_{\alpha}$.
It turs out that the factors $D_{L}({\bf Q},{\bf P})$ and $D_{R}({\bf Q},{\bf P})$, eqs(\ref{30})-(\ref{31}),
can be nicely represented in the determinant form \cite{Gauden}:
\begin{equation}
 \label{34}
D_{L}({\bf Q},{\bf P}) = \delta_{L}^{-1}({\bf Q},{\bf P}) \;
\det\Bigl[\frac{1}{(Q_{a}-P_{b}-i\epsilon)(Q_{a} - P_{b}-i\epsilon + i\kappa)}\Bigr]_{a,b = 1, .., L}
\end{equation}
where
\begin{equation}
 \label{35}
\delta_{L}({\bf Q},{\bf P}) = \det\Bigl[\frac{1}{Q_{a} - P_{b}-i\epsilon + i\kappa}\Bigr]_{a,b = 1, .., L}
= \frac{\prod_{1\leq a<b}^{L}\bigl[(Q_{a}-Q_{b})(P_{a}-P_{b})\bigr]}{
\prod_{a,b=1}^{L}\bigl(Q_{a}-P_{b}-i\epsilon+i\kappa\bigr)}
\end{equation}
The expression for $D_{R}({\bf Q},{\bf P})$ is obtained from eqs.(\ref{34})-(\ref{35}) by changing
$Q_{a} \to Q^{*}_{a}$, $P_{a} \to P^{*}_{a}$ and $L \to R$.

Thus, substituting eqs.(\ref{29}), (\ref{10}) and (\ref{11}) into eq.(\ref{24}) and taking into account
that the wave function $\Psi_{\bf P}({\bf x}, {\bf y})$ exists only provided $(m_{\alpha}+s_{\alpha}) \geq 1$
we get:
\begin{eqnarray}
 \nonumber
W(f_{1},f_{2},\Delta) &=&
1 + \lim_{t\to\infty}\sum_{M=1}^{\infty} \frac{(-1)^{M}}{M!}
\prod_{\alpha=1}^{M}
\Biggl[
\int_{-\infty}^{+\infty} \frac{dq_{\alpha}}{2\pi}
\Biggr]
\sum_{M_{1}=0}^{M}
\frac{1}{M_{1}!}
\prod_{\beta=1}^{M_{1}}
\Biggl[
\int_{-\infty}^{+\infty} \frac{dp_{\beta}}{2\pi}
\Biggr]
\times
\\
\nonumber
\\
\nonumber
&\times&
\frac{M!}{M_{1}!(M-M_{1})!} \;
\prod_{\beta=1}^{M_{1}}
\Biggl[
\sum_{k_{\beta}=0}^{\infty} \sum_{m_{\beta}+s_{\beta}\geq 1}^{\infty} (-1)^{m_{\beta}+k_{\beta}+s_{\beta}-1}\Biggr]
\prod_{\gamma=M_{1}+1}^{M}
\Biggl[
\sum_{k_{\gamma}=1}^{\infty} \delta_{m_{\gamma},0} \; \delta_{s_{\gamma},0} (-1)^{k_{\gamma}-1}\Biggr]
\times
\\
\nonumber
\\
\nonumber
&\times&
\exp
\Biggl\{
\sum_{\alpha=1}^{M}
\Bigl[
\lambda_{t} k_{\alpha} f_{1} + (1+\Delta)^{1/3} \lambda_{t} (m_{\alpha}+s_{\alpha})f_{2}
- t E(q_{\alpha}, m_{\alpha}+k_{\alpha}+s_{\alpha})
\Bigr]
- \Delta \, t \sum_{\beta=1}^{M_{1}} E(p_{\beta}, m_{\beta}+s_{\beta})
\Biggr\}
\times
\\
\nonumber
\\
\nonumber
&\times&
\det\Biggl[
\frac{1}{
\frac{1}{2}\kappa( m_{\alpha}+k_{\alpha}+s_{\alpha}) -iq_{\alpha} +
\frac{1}{2}\kappa( m_{\alpha'}+k_{\alpha'}+s_{\alpha'}) +iq_{\alpha'}}
\Biggr]_{\alpha,\alpha'=1,...,M} \;
\times
\\
\nonumber
\\
\nonumber
&\times&
\det\Biggl[
\frac{1}{
\frac{1}{2}\kappa( m_{\beta}+s_{\beta}) +ip_{\beta} +
\frac{1}{2}\kappa( m_{\beta'}+s_{\beta'}) -ip_{\beta'}}
\Biggr]_{\beta,\beta'=1,...,M_{1}} \;
\times
\\
\nonumber
\\
&\times&
D_{L} \bigl({\bf q}, {\bf p}, {\bf m}, {\bf k}, {\bf s}\bigr) \;
D_{R} \bigl({\bf q}, {\bf p}, {\bf m}, {\bf k}, {\bf s}\bigr) \;
{\cal G} \bigl({\bf q}, {\bf p}, {\bf m}, {\bf k}, {\bf s}\bigr)
\label{36}
\end{eqnarray}
where
\begin{equation}
 \label{37}
 E(q,n) \; = \; \frac{1}{2\beta} n q^{2} - \frac{\kappa^{2}}{24\beta} n^{3}
\end{equation}
and instead of the vectors ${\bf Q}$ and ${\bf P}$ in the arguments of the functions $D_{L}$, $D_{R}$ and ${\cal G}$
we have introduced the vectors ${\bf q} = \{q_{1}, ..., q_{M}\}$, ${\bf p} = \{p_{1}, ..., p_{M_{1}}\}$,
${\bf m} = \{m_{1}, ..., m_{M}\}$, ${\bf k} = \{k_{1}, ..., k_{M}\}$ and ${\bf s} = \{s_{1}, ..., s_{M}\}$.

\vspace{5mm}

Now using explicit expressions for the momenta components $q^{\alpha}_{r}$ and $p^{\alpha}_{r}$,
eqs.(\ref{32})-(\ref{33}), we have to express the factors $D_{L}$, $D_{R}$, eqs.(\ref{34})-(\ref{35}),
and ${\cal G}$, eq.(\ref{38}), in terms of analytic functions of the parameters
$q_{\alpha}, p_{\alpha}, m_{\alpha}, k_{\alpha}$ and $s_{\alpha}$.
One can easily note that due to the symmetry of the expressions (\ref{36}) and (\ref{34})-(\ref{35})
with respect to permutations of the clusters, the factors $D_{L}$ and  $D_{R}$
in eq.(\ref{36}) can be represented as follows:
\begin{eqnarray}
 \label{42}
D_{L} &=& \frac{L!}{\prod_{\alpha=1}^{M} m_{\alpha}! } \; \delta_{L}^{-1} \; \prod_{\alpha=1}^{M}
d_{\alpha}^{(L)}
\\
\nonumber
\\
D_{R} &=& \frac{R!}{\prod_{\alpha=1}^{M} s_{\alpha}! } \; \delta_{R}^{-1} \; \prod_{\alpha=1}^{M}
d_{\alpha}^{(R)}
\label{43}
\end{eqnarray}
where
\begin{eqnarray}
 \label{44}
d_{\alpha}^{(L)} &=&
\det\Bigl[
\frac{1}{
\bigl(q^{\alpha}_{r} - p^{\alpha}_{r'}-i\epsilon\bigr)
\bigl(q^{\alpha}_{r} - p^{\alpha}_{r'}-i\epsilon + i\kappa\bigr)}
\Bigr]_{r,r'=1,...,m_{\alpha}}
\\
\nonumber
\\
d_{\alpha}^{(R)} &=&
\det\Bigl[
\frac{1}{
\bigl({q^{*}}^{\alpha}_{r} - {p^{*}}^{\alpha}_{r'}+i\epsilon\bigr)
\bigl({q^{*}}^{\alpha}_{r} - {p^{*}}^{\alpha}_{r'}+i\epsilon - i\kappa\bigr)}
\Bigr]_{r,r'=1,...,s_{\alpha}}
 \label{45}
\end{eqnarray}
Using  eq.(\ref{32}) the determinant (\ref{44}) can be represented as follows:
\begin{equation}
 \label{46a}
d_{\alpha}^{(L)} \; =  \; (i\kappa)^{-m_{\alpha}} \;
\det\Bigl[
\frac{1}{
(x_{\alpha}+r+r'-2)(x_{\alpha}+r+r'-1)}
\Bigr]_{r,r'=1,...,m_{\alpha}}
\end{equation}
where
\begin{equation}
 \label{47}
x_{\alpha} = -i A_{\alpha}^{(-)} - m_{\alpha} - s_{\alpha} - \frac{1}{2} k_{\alpha} + 1
\end{equation}
and
\begin{equation}
 \label{48}
A_{\alpha}^{(-)} = \frac{1}{\kappa} (q_{\alpha}-p_{\alpha}) - i\epsilon
\end{equation}
Sufficiently simple calculations yield:
\begin{equation}
 \label{46}
d_{\alpha}^{(L)} \; = \;
(i\kappa)^{-m_{\alpha}} \;
\frac{m_{\alpha}! \prod_{l=1}^{m_{\alpha}} l^{2(m_{\alpha}-l)}}{
\prod_{l=1}^{m_{\alpha}}\Bigl[(x_{\alpha} +l -1)^{l} (x_{\alpha} + 2m_{\alpha} - l)^{l}\Bigr]}
\end{equation}
The expression for $D_{\alpha}^{(R)}$ is obtained from eqs.(\ref{44}) and (\ref{46})
by changing $m_{\alpha} \to s_{\alpha}$ and $i A_{\alpha}^{(-)} \to -i A_{\alpha}^{(+)}$.

After somewhat painful algebra for the factor $\delta_{L}^{-1}$ one obtains the following expression:
\begin{equation}
 \label{49}
\delta_{L}^{-1} = (i\kappa)^{L}
\frac{\prod_{\alpha=1}^{M} \prod_{r,r'=1}^{m_{\alpha}}
\Bigl[-i A_{\alpha}^{(-)} - n_{\alpha} -\frac{1}{2}k_{\alpha} +r +r'\Bigr] \;
\prod_{\alpha\not=\beta}^{M} \prod_{r=1}^{m_{\alpha}}\prod_{r'=1}^{m_{\beta}}
\Bigl[R_{\alpha\beta} +r +r'\Bigr]}{
\prod_{\alpha=1}^{M} \prod_{1\leq r<r'}^{m_{\alpha}} (r-r')^{2} \;
\prod_{1\leq\alpha<\beta}^{M} \prod_{r=1}^{m_{\alpha}}\prod_{r'=1}^{m_{\beta}}
\Bigl[\bigl(Q_{\alpha\beta} +r-r'\bigr)\bigl(P_{\alpha\beta} +r-r'\bigr)\Bigr]}
\end{equation}
where
\begin{eqnarray}
 \nonumber
Q_{\alpha\beta} &=& -\frac{i}{\kappa} (q_{\alpha}-q_{\beta}) - \frac{1}{2}(n_{\alpha}'-n_{\beta}')
\\
\nonumber
\\
\label{50}
P_{\alpha\beta} &=& \frac{i}{\kappa} (p_{\alpha}-p_{\beta}) - \frac{1}{2}(n_{\alpha}-n_{\beta})
\\
\nonumber
\\
\nonumber
R_{\alpha\beta} &=& -\frac{i}{\kappa} (q_{\alpha}-p_{\beta}) - \frac{1}{2}(n_{\alpha}'-n_{\beta})
\end{eqnarray}
and
\begin{eqnarray}
 \nonumber
n_{\alpha} &=& m_{\alpha} + s_{\alpha}
\\
\label{51}
\\
\nonumber
n_{\alpha}' &=& m_{\alpha} + s_{\alpha} +k_{\alpha}
\end{eqnarray}
Using the properties of the Barnes $G$-function:
\begin{eqnarray}
 \nonumber
G(N+1) &=& \prod_{l=1}^{N} \Gamma(l)
\\
\nonumber
\\
\nonumber
G(z+1) &=& \Gamma(z) G(z)
\\
\label{52}
\\
\nonumber
\prod_{l=1}^{N} \Gamma(z+l) &=& \frac{G(z+N+1)}{G(z+1)}
\\
\nonumber
\\
\nonumber
\prod_{l=1}^{N} (z+l)^{l} &=& \Gamma^{N}(z+N+1) \, \frac{G(z+1)}{G(z+N+1)}
\end{eqnarray}
for the factors (\ref{46}) and (\ref{49}) one eventually obtains the following expressions:
\begin{eqnarray}
\label{53}
d_{\alpha}^{(L)} &=& (i\kappa)^{-m_{\alpha}} \;
\frac{\Gamma(m_{\alpha} +1) G^{2}(m_{\alpha}+1) G(n_{\alpha} + \frac{1}{2}k_{\alpha} - m_{\alpha} + i A_{\alpha}^{(-)} +1)}{
G(n_{\alpha} + \frac{1}{2}k_{\alpha} + i A_{\alpha}^{(-)} +1)}
\times
\frac{G(n_{\alpha} + \frac{1}{2}k_{\alpha} - m_{\alpha} + i A_{\alpha}^{(-)})}{
G(n_{\alpha} + \frac{1}{2}k_{\alpha} - 2m_{\alpha} + i A_{\alpha}^{(-)})}
\\
\nonumber
\\
\delta_{L}^{-1} &=& \prod_{\alpha=1}^{M}
\Biggl[
\frac{(i\kappa)^{-m_{\alpha}} G(n_{\alpha} + \frac{1}{2}k_{\alpha} + i A_{\alpha}^{(-)})
G(n_{\alpha} + \frac{1}{2}k_{\alpha} - 2m_{\alpha} + i A_{\alpha}^{(-)})}{
G^{2}(m_{\alpha}+1) G^{2}(n_{\alpha} + \frac{1}{2}k_{\alpha} - m_{\alpha} + i A_{\alpha}^{(-)})}
\Biggr]
\times
\prod_{\alpha\not=\beta}^{M} {\cal B}_{\alpha\beta}^{(L)}
\label{54}
\end{eqnarray}
where
\begin{eqnarray}
 \label{55}
{\cal B}_{\alpha\beta}^{(L)} &=&
\frac{G(Q_{\alpha\beta} + 1) G(P_{\alpha\beta}+1)
G(Q_{\alpha\beta} + m_{\alpha}-m_{\beta} +1)
G(P_{\alpha\beta} - m_{\alpha}+m_{\beta} +1)}{
G(Q_{\alpha\beta} + m_{\alpha} +1)G(P_{\alpha\beta} - m_{\alpha} +1)
G(Q_{\alpha\beta} - m_{\beta} +1)G(P_{\alpha\beta} + m_{\beta} +1)}
\times
\\
\nonumber
\\
\nonumber
&\times&
\frac{G(R_{\alpha\beta} + m_{\alpha}+m_{\beta} +2)
G(R_{\beta\alpha} + m_{\alpha}+m_{\beta} +2)
G(R_{\alpha\beta} +2)G(R_{\beta\alpha} +2)}{
G(R_{\alpha\beta} + m_{\alpha} +2)G(R_{\beta\alpha}+m_{\beta} +2)
G(R_{\alpha\beta} + m_{\beta} +2) G(R_{\beta\alpha} + m_{\alpha} +2)}
\end{eqnarray}
Finally, substituting eqs.(\ref{53}) and (\ref{54}) into eq.(\ref{42}) we get:
\begin{equation}
 \label{56}
D_{L} = \Gamma\bigl(\sum_{\alpha}m_{\alpha} +1\bigr)
\prod_{\alpha=1}^{M}
\Biggl[
\frac{G(n_{\alpha} + \frac{1}{2}k_{\alpha} - m_{\alpha} + i A_{\alpha}^{(-)} +1)
G(n_{\alpha} + \frac{1}{2}k_{\alpha}  + i A_{\alpha}^{(-)})}{
G(n_{\alpha} + \frac{1}{2}k_{\alpha}  + i A_{\alpha}^{(-)} +1)
G(n_{\alpha} + \frac{1}{2}k_{\alpha} - m_{\alpha} + i A_{\alpha}^{(-)})}
\Biggr]
\times
\prod_{\alpha\not=\beta}^{M} {\cal B}_{\alpha\beta}^{(L)}
\end{equation}
The expression for $D_{R}$ is obtained from eq.(\ref{56}) by changing
$m_{\alpha} \to s_{\alpha}$ and $iA_{\alpha}^{(-)} \to -iA_{\alpha}^{(+)}$.
In the similar way one can derive the analytic expression for the cross-product factor
${\cal G}$, eq.(\ref{38}). The final formula for this factor is rather cumbersome:
it contains the  products of all kinds of Gamma functions of the type
$\Gamma\bigl[1 + \frac{1}{2}(\pm m_{\alpha}\pm s_{\alpha}\pm k_{\alpha}
\pm m_{\beta}\pm s_{\beta}\pm k_{\beta}) \pm i (q_{\alpha} - q_{\beta})/\kappa\bigr]$
and $\Gamma\bigl[1 + \frac{1}{2}(\pm m_{\alpha}\pm s_{\alpha}
\pm m_{\beta}\pm s_{\beta}) \pm i (p_{\alpha} - p_{\beta})/\kappa\bigr]$
(the example of such type of product one can see in \cite{end-point}, eq.(A.17)).
We do not reproduce it here as it turns out to be irrelevant in the limit $t\to\infty$
(see below).

\subsection{Thermodynamic limit $t \to \infty$}

Next step of the calculations is to take the limit $t \to \infty$ and to perform the
summations over the integers $m_{\alpha}, s_{\alpha}$ and $k_{\alpha}$
After rescaling
\begin{eqnarray}
 \nonumber
q_{\alpha} &\to& \frac{\kappa}{2\lambda_{t}} \, q_{\alpha}
\\
\label{57}
\\
\nonumber
p_{\alpha} &\to& \frac{\kappa}{2\lambda_{t}} \, p_{\alpha}
\end{eqnarray}
with
\begin{equation}
 \label{58}
\lambda_{t} \; = \; \frac{1}{2} \Bigl(\frac{\kappa^{2} t}{\beta}\Bigr)^{1/3} \; = \;
\frac{1}{2} \bigl(\beta^{5} u^{2} t\bigr)^{1/3}
\end{equation}
the expression for the probability distribution function
$W(f_{1}, f_{2}, \Delta)$, eq.(\ref{36}), reduces to
\begin{eqnarray}
 \nonumber
W(f_{1},f_{2},\Delta) &=&
1 + \lim_{\lambda_{t}\to\infty}
\sum_{M=1}^{\infty} \sum_{M_{1}=1}^{M}
\frac{(-1)^{M}}{(M_{1}!)^{2}(M-M_{1})!}
\times
\\
\nonumber
\\
\nonumber
&\times&
\prod_{\beta=1}^{M_{1}}
\Biggl[
\int\int_{-\infty}^{+\infty} \frac{dq_{\beta}dp_{\beta}}{(2\pi)^{2}}
\sum_{k_{\beta}=0}^{\infty} \sum_{m_{\beta}+s_{\beta}\geq 1}^{\infty} (-1)^{m_{\beta}+k_{\beta}+s_{\beta}-1}
\times
\\
\nonumber
\\
\nonumber
&\times&
\exp
\Bigl\{
\lambda_{t} k_{\beta} f_{1} + (1+\Delta)^{1/3} \lambda_{t} (m_{\beta}+s_{\beta})f_{2}
- \lambda (m_{\beta}+k_{\beta}+s_{\beta}) q_{\beta}^{2}
+ \frac{1}{3}\lambda_{t}^{3} (m_{\beta}+k_{\beta}+s_{\beta})^{3}
\\
\nonumber
\\
\nonumber
&-& \Delta \, \lambda_{t} (m_{\beta}+s_{\beta}) p_{\beta}^{2}
+ \frac{1}{3} \Delta \, \lambda_{t}^{3} (m_{\beta}+s_{\beta})^{3}
\Bigr\} \;
J\Bigl(\frac{q_{\beta}-p_{\beta}}{\lambda_{t}},  m_{\beta}, s_{\beta}, k_{\beta}\Bigr)
\Biggr]
\times
\\
\nonumber
\\
\nonumber
&\times&
\prod_{\gamma=M_{1}+1}^{M}
\Biggl[
\int_{-\infty}^{+\infty} \frac{dq_{\gamma}}{2\pi}
\sum_{k_{\gamma}=1}^{\infty} \delta_{m_{\gamma},0} \; \delta_{s_{\gamma},0} (-1)^{k_{\gamma}-1}
\exp
\Bigl\{
\lambda_{t} k_{\gamma} f_{1} -  \lambda_{t} k_{\gamma} q_{\gamma}^{2} + \frac{1}{3}\lambda_{t}^{3} k_{\gamma}^{3}
\Bigr\}
\Biggr]
\times
\\
\nonumber
\\
\nonumber
&\times&
\det\Biggl[
\frac{1}{
\lambda_{t}( m_{\alpha}+k_{\alpha}+s_{\alpha}) -iq_{\alpha} +
\lambda_{t}( m_{\alpha'}+k_{\alpha'}+s_{\alpha'}) +iq_{\alpha'}}
\Biggr]_{\alpha,\alpha'=1,...,M} \;
\times
\\
\nonumber
\\
\nonumber
&\times&
\det\Biggl[
\frac{1}{
\lambda_{t}( m_{\beta}+s_{\beta}) +ip_{\beta} +
\lambda_{t}( m_{\beta'}+s_{\beta'}) -ip_{\beta'}}
\Biggr]_{\beta,\beta'=1,...,M_{1}} \;
\times
\\
\nonumber
\\
&\times&
\prod_{\alpha\not=\alpha'}^{M}\Bigl[ {\cal B}_{\alpha\alpha'}^{(L)}{\cal B}_{\alpha\alpha'}^{(R)}\Bigr] \;
{\cal G}
\Biggr\}
\label{59}
\end{eqnarray}
Here
\begin{eqnarray}
 \nonumber
J\Bigl(\frac{q-p}{\lambda_{t}}, m, s, k\Bigr)
&=&
\Gamma(m +1) \Gamma(s + 1)
\times
\\
\nonumber
\\
\nonumber
&\times&
\frac{G\bigl(n + \frac{1}{2}k - m + \frac{i}{2\lambda_{t}} (q-p)^{(-)} +1\bigr)
      G\bigl(n + \frac{1}{2}k - s - \frac{i}{2\lambda_{t}} (q-p)^{(+)} +1\bigr)}{
      G\bigl(n + \frac{1}{2}k  + \frac{i}{2\lambda_{t}} (q-p)^{(-)} +1\bigr)
      G\bigl(n + \frac{1}{2}k  - \frac{i}{2\lambda_{t}} (q-p)^{(+)} +1\bigr)}
\times
\\
\nonumber
\\
&\times&
\frac{G\bigl(n + \frac{1}{2}k  + \frac{i}{2\lambda_{t}} (q-p)^{(-)}\bigr)
      G\bigl(n + \frac{1}{2}k  - \frac{i}{2\lambda_{t}} (q-p)^{(+)}\bigr)}{
      G\bigl(n + \frac{1}{2}k - m + \frac{i}{2\lambda_{t}} (q-p)^{(-)}\bigr)
      G\bigl(n + \frac{1}{2}k - s - \frac{i}{2\lambda_{t}} (q-p)^{(+)}\bigr)}
\label{60}
\end{eqnarray}
where $n = m + s$ and $(q-p)^{(\pm)} \; \equiv \; q - p \pm i \epsilon$.

The summations over the integers $m_{\alpha}, s_{\alpha}$ and $k_{\alpha}$ in the limit
$\lambda\to\infty$ is performed according to the following heuristic algorithm
(see also \cite{goe, end-point}).
Let us consider the example of the series of a general type:
\begin{equation}
\label{61}
R = \lim_{\lambda\to\infty} \;
\prod_{\alpha=1}^{M}\Biggl[\sum_{n_{\alpha}=0}^{\infty} (-1)^{n_{\alpha}-1}\Biggr] \;
\Phi \bigl(\lambda; \,  \lambda {\bf n}; \, {\bf n}\bigr)
\end{equation}
where $\Phi \bigl(\lambda; \,  \lambda z_{1}, ..., \lambda z_{M}; \, z_{1}, ..., z_{M}\bigr)$
as a function of the variables $\{z_{1}, ..., z_{M}\}$ has "good" analytic properties in
the complex half-plane, $\mbox{Re}(z_{\alpha}) \geq 0$. 
Then the summations in the above equation can be represented in
terms of the following contour integrals:
\begin{equation}
\label{62}
R = \lim_{\lambda\to\infty} \;
\prod_{\alpha=1}^{M}
\Biggl[
\frac{1}{2i} \int_{{\cal C}} \frac{dz_{\alpha}}{\sin(\pi z_{\alpha})}
\Biggr] \;
\Phi \bigl(\lambda; \,  \lambda {\bf z}; \, {\bf z}\bigr)
\end{equation}
where the integration contour ${\cal C}$ is shown in Figure 2a.
After rescaling, $z_{\alpha} \to z_{\alpha}/\lambda$ we get:
\begin{equation}
\label{63}
R =  \;
\prod_{\alpha=1}^{M}
\Biggl[
\frac{1}{2\pi i} \int_{{\cal C}} \frac{dz_{\alpha}}{z_{\alpha}}
\Biggr] \;
\lim_{\lambda\to\infty} \;
\Phi \bigl(\lambda; \,  {\bf z}; \, {\bf z}/\lambda \bigr)
\end{equation}
where the parameters $z_{\alpha}$ remain finite in the limit $\lambda \to \infty$.
Note that if the summation in eq.(\ref{61}) starts at $n=1$ (and not at $n=0$),
the integration in eq.(\ref{63}) goes along the contour ${\cal C}'$ shown in Figure 2b.

\begin{figure}[h]
\begin{center}
   \includegraphics[width=12.0cm]{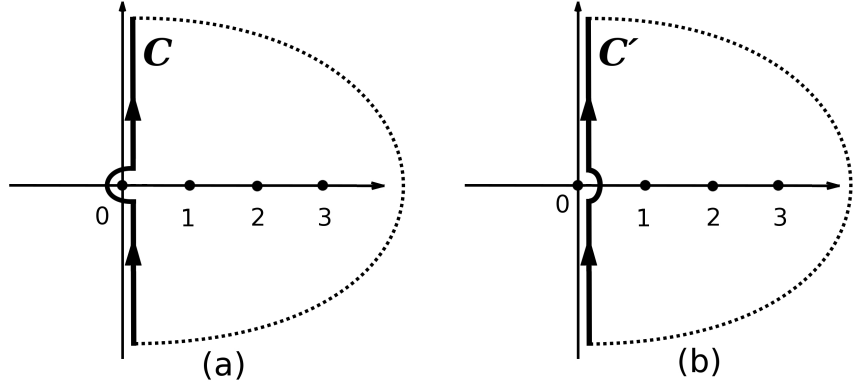}
\caption[]{
(a) Contour of integration ${\cal C}$ in eq.(\ref{62});
(b) Contour of integration ${\cal C}'$ in the case the summation of the series (\ref{61})
starts at $n=1$}
\end{center}
\label{figure2}
\end{figure}

Applying the above algorithm for the expression in eq.(\ref{59}) and taking into account that
\begin{eqnarray}
\nonumber
\lim_{|z|\to 0} \; \Gamma(z+1) &=& 1
\\
\nonumber
\\
\label{64}
\lim_{|z|\to 0} \; G(z+1) &=& 1
\\
\nonumber
\\
\nonumber
G(z)\big|_{|z|\ll 1} &\simeq& z
\end{eqnarray}
we find:
\begin{eqnarray}
\nonumber
\lim_{\lambda_{t}\to\infty} \;
{\cal B}_{\alpha\beta}^{(L,R)} &=& 1
\\
\label{65}
\\
\nonumber
\lim_{\lambda_{t}\to\infty} \;
{\cal G} &=& 1
\end{eqnarray}
and
\begin{equation}
\label{66}
\lim_{\lambda_{t}\to\infty} \;
J\Bigl(\frac{q-p}{\lambda_{t}},  \frac{z_{1}}{\lambda_{t}}, \frac{z_{2}}{\lambda_{t}}, \frac{z_{3}}{\lambda_{t}}\Bigr)
=
\Biggl(1 + \frac{z_{1}}{z_{2} + \frac{1}{2} z_{3} + \frac{1}{2} i (q-p)^{(-)}}\Biggr)
\Biggl(1 + \frac{z_{2}}{z_{1} + \frac{1}{2} z_{3} - \frac{1}{2} i (q-p)^{(+)}}\Biggr)
\equiv
J_{*}\bigl(q-p,z_{1},z_{2},z_{3}\bigr)
\end{equation}
Thus, instead of eq.(\ref{59}), in the limit $\lambda_{t}\to\infty$ we get the following
much more simple expression:
\begin{eqnarray}
 \nonumber
W(f_{1},f_{2},\Delta) &=&
1 + \sum_{M=1}^{\infty} \sum_{M_{1}=1}^{M}
\frac{(-1)^{M}}{(M_{1}!)^{2}(M-M_{1})!}
\\
\nonumber
\\
\nonumber
&\times&
\prod_{\beta=1}^{M_{1}}
\Biggl[
\int\int_{-\infty}^{+\infty} \frac{dq_{\beta}dp_{\beta}}{(2\pi)^{2}}
\int\int_{-\infty}^{+\infty} \frac{dy_{1} dy_{2}}{\Delta^{1/3}}
\Ai\bigl(y_{1} + q_{\beta}^{2} - f_{1}\bigr)
\Ai\Bigl(\frac{y_{2} + \Delta \, p_{\beta}^{2} - (1+\Delta)^{1/3} f_{2} + f_{1}}{\Delta^{1/3}}\Bigr)
\times
\\
\nonumber
\\
\nonumber
&\times&
\int_{{\cal C}} \frac{d{z_{3}}_{\beta}}{{z_{3}}_{\beta}}
\int_{{\cal C}} d{z_{1}}_{\beta}
\int_{{\cal C}} d{z_{2}}_{\beta}
\Bigl(\frac{1}{{z_{1}}_{\beta} {z_{2}}_{\beta}} - \delta({z_{1}}_{\beta}) \delta({z_{2}}_{\beta}) \Bigr)
J_{*}\Bigl(q_{\beta}- p_{\beta}, {z_{1}}_{\beta}, {z_{2}}_{\beta}, {z_{3}}_{\beta}\Bigr)
\times
\\
\nonumber
\\
\nonumber
&\times&
\exp\bigl\{ ({z_{1}}_{\beta}+ {z_{2}}_{\beta})(y_{1} + y_{2}) + {z_{3}}_{\beta} y_{1}\bigr\}
\Biggr]
\times
\\
\nonumber
\\
\nonumber
&\times&
\prod_{\gamma=M_{1}+1}^{M}
\Biggl[
\int\int_{-\infty}^{+\infty} \frac{dq_{\gamma}dy_{1}}{2\pi}
\Ai\bigl(y_{1} + q_{\gamma}^{2} - f_{1}\bigr)
\int_{{\cal C}'} \frac{d{z_{3}}_{\gamma}}{{z_{3}}_{\gamma}}
\int_{{\cal C}} d{z_{1}}_{\gamma}
\int_{{\cal C}} d{z_{2}}_{\gamma}
\delta({z_{1}}_{\gamma}) \delta({z_{2}}_{\gamma})
\exp\bigl\{ {z_{3}}_{\gamma} y_{1} \bigr\}
\Biggr]
\times
\\
\nonumber
\\
\nonumber
&\times&
\det\Biggl[
\frac{1}{
{z_{1}}_{\alpha} + {z_{2}}_{\alpha} + {z_{3}}_{\alpha}-iq_{\alpha} +
{z_{1}}_{\alpha'} + {z_{2}}_{\alpha'} + {z_{3}}_{\alpha'}+iq_{\alpha'}}
\Biggr]_{\alpha,\alpha'=1,...,M}
\times
\\
\nonumber
\\
&\times&
\det\Biggl[
\frac{1}{
{z_{1}}_{\beta} + {z_{2}}_{\beta} +ip_{\beta} +
{z_{1}}_{\beta'} + {z_{2}}_{\beta'} -ip_{\beta'}}
\Biggr]_{\beta,\beta'=1,...,M_{1}}
\label{67}
\end{eqnarray}
where we have  used the Airy function identity
\begin{equation}
\label{69}
\exp\Bigl\{\frac{1}{3} z^{3} \Bigr\} \; = \;
\int_{-\infty}^{+\infty} dy \; \Ai(y) \; \exp\bigl\{zy\bigr\}
\end{equation}

\subsection{The Result}

One can easily see that the expression for the function $W(f_{1},f_{2},\Delta)$,
eq.(\ref{67}), is {\it not} the Fredholm determinant. Consequently, it can not be represented
in the exponential form in terms of the trace of the corresponding matrix which was the
standard step in all previous  Bethe ansatz replica calculations of the Fredholm determinant integral
kernels (see e.g. \cite{BA-TW2, LeDoussal1, rev-TW}). Nevertheless, sufficiently simple
structure of two determinants in eq.(\ref{67}) allows to perform significant simplification
of this expression.

Indeed, for the first $M\times M$ determinant in eq.(\ref{67}) we have:
\begin{eqnarray}
 \nonumber
\mbox{det}_{M} &\equiv&
\det\Biggl[
\frac{1}{
{z_{1}}_{\alpha} + {z_{2}}_{\alpha} + {z_{3}}_{\alpha}-iq_{\alpha} +
{z_{1}}_{\alpha'} + {z_{2}}_{\alpha'} + {z_{3}}_{\alpha'}+iq_{\alpha'}}
\Biggr]_{\alpha,\alpha'=1,...,M}
\\
\nonumber
\\
\nonumber
&=&
\sum_{{\cal P}\in S_{M}} (-1)^{\bigl[{\cal P}\bigr]}
\prod_{\alpha=1}^{M}
\bigl(
{z_{1}}_{\alpha} + {z_{2}}_{\alpha} + {z_{3}}_{\alpha} - iq_{\alpha} +
{z_{1}}_{{\cal P}_\alpha} + {z_{2}}_{{\cal P}_\alpha} + {z_{3}}_{{\cal P}_\alpha}+iq_{{\cal P}_\alpha}
\bigr)^{-1}
\\
\nonumber
\\
\nonumber
&=&
\prod_{\alpha=1}^{M}
\Biggl[
\int_{0}^{\infty} du_{\alpha}
\Biggr]
\sum_{{\cal P}\in S_{M}} (-1)^{\bigl[{\cal P}\bigr]}
\exp
\Bigl\{
-\sum_{\alpha=1}^{M}
\bigl(
{z_{1}}_{\alpha} + {z_{2}}_{\alpha} + {z_{3}}_{\alpha}
\bigr) u_{\alpha}
-\sum_{\alpha=1}^{M}
\bigl(
{z_{1}}_{{\cal P}_\alpha} + {z_{2}}_{{\cal P}_\alpha} + {z_{3}}_{{\cal P}_\alpha}
\bigr) u_{\alpha}
\Bigr\}
\times
\\
\nonumber
\\
&\times&
\exp
\Bigl\{
-\sum_{\alpha=1}^{M}
q_{\alpha} u_{\alpha}
-\sum_{\alpha=1}^{M}
q_{{\cal P}_\alpha} u_{\alpha}
\Bigr\}
\label{70}
\end{eqnarray}
Since
\begin{equation}
\label{71}
\sum_{\alpha=1}^{M}
\bigl(
{z_{1}}_{{\cal P}_\alpha} + {z_{2}}_{{\cal P}_\alpha} + {z_{3}}_{{\cal P}_\alpha}
\bigr)u_{\alpha}
\; = \;
\sum_{\alpha=1}^{M}
\bigl(
{z_{1}}_{\alpha} + {z_{2}}_{\alpha} + {z_{3}}_{\alpha}
\bigr) u_{{\cal P}^{-1}_\alpha}
\end{equation}
and
\begin{equation}
\label{72}
\sum_{\alpha=1}^{M}
q_{{\cal P}_\alpha} u_{\alpha}
\; = \;
\sum_{\alpha=1}^{M}
q_{\alpha} u_{{\cal P}^{-1}_\alpha}
\end{equation}
redefining ${\cal P}^{-1} \to {\cal P}$ we get
\begin{equation}
\label{73}
\mbox{det}_{M} =
\prod_{\alpha=1}^{M}
\Biggl[
\int_{0}^{\infty} du_{\alpha}
\Biggr]
\sum_{{\cal P}\in S_{M}} (-1)^{\bigl[{\cal P}\bigr]}
\prod_{\alpha=1}^{M}
\exp
\Biggl\{
-\sum_{\alpha=1}^{M}
\bigl(
{z_{1}}_{\alpha} + {z_{2}}_{\alpha} + {z_{3}}_{\alpha}
\bigr) \bigl(u_{\alpha} + u_{{\cal P}_\alpha}\bigr)
+ i \sum_{\alpha=1}^{M}
q_{\alpha} \bigl(u_{\alpha} - u_{{\cal P}_\alpha}\bigr)
\Biggr\}
\end{equation}
In the similar way for the second determinant in eq.(\ref{67}) we obtain
\begin{equation}
\label{74}
\mbox{det}_{M_{1}} =
\prod_{\beta=1}^{M_{1}}
\Biggl[
\int_{0}^{\infty} dv_{\beta}
\Biggr]
\sum_{\tilde{{\cal P}}\in S_{M_{1}}} (-1)^{\bigl[\tilde{{\cal P}}\bigr]}
\prod_{\beta=1}^{M_{1}}
\exp
\Biggl\{
-\sum_{\beta=1}^{M_{1}}
\bigl(
{z_{1}}_{\beta} + {z_{2}}_{\beta}
\bigr) \bigl(v_{\beta} + v_{\tilde{{\cal P}}_\beta}\bigr)
- i \sum_{\beta=1}^{M_{1}}
p_{\beta} \bigl(v_{\beta} - v_{\tilde{{\cal P}}_\beta}\bigr)
\Biggr\}
\end{equation}
Substituting eqs.(\ref{73})-(\ref{74}) into eq.(\ref{67}) we get
\begin{eqnarray}
 \nonumber
W(f_{1},f_{2},\Delta) &=&
1 + \sum_{M=1}^{\infty} \sum_{M_{1}=1}^{M}
\frac{(-1)^{M}}{(M_{1}!)^{2}(M-M_{1})!}
\prod_{\alpha=1}^{M}
\Biggl[
\int_{0}^{\infty} du_{\alpha}
\Biggr]
\prod_{\beta=1}^{M_{1}}
\Biggl[
\int_{0}^{\infty} dv_{\beta}
\Biggr]
\sum_{{\cal P}\in S_{M}} (-1)^{\bigl[{\cal P}\bigr]}
\sum_{\tilde{{\cal P}}\in S_{M_{1}}} (-1)^{\bigl[\tilde{{\cal P}}\bigr]}
\\
\nonumber
\\
\nonumber
&\times&
\prod_{\beta=1}^{M_{1}}
\Biggl[
\int\int_{-\infty}^{+\infty} \frac{dqdp}{(2\pi)^{2}}
\int\int_{-\infty}^{+\infty} \frac{dy_{1} dy_{2}}{\Delta^{1/3}} \;
\Ai\bigl(y_{1} + q^{2} - f_{1}\bigr)
\;
\exp\bigl\{i q \bigl(u_{\beta} - u_{{\cal P}_{\beta}}\bigr)\bigr\}
\times
\\
\nonumber
\\
\nonumber
&\times&
\Ai\Bigl(\frac{y_{2} + \Delta \, p^{2} - (1+\Delta)^{1/3} f_{2} + f_{1}}{\Delta^{1/3}}\Bigr)
\exp\bigl\{-i p \bigl(v_{\beta} - v_{\tilde{{\cal P}}_{\beta}}\bigr)\bigr\}
\times
\\
\nonumber
\\
\nonumber
&\times&
\int_{{\cal C}} \frac{dz_{3}}{z_{3}}
\int_{{\cal C}} dz_{1}
\int_{{\cal C}} dz_{2}
\Bigl(\frac{1}{z_{1} z_{2}} - \delta(z_{1}) \delta(z_{2}) \Bigr) \;
J_{*}\Bigl(q- p, z_{1}, z_{2}, z_{3}\Bigr)
\times
\\
\nonumber
\\
\nonumber
&\times&
\exp\Bigl\{
\bigl(z_{1} + z_{2}\bigr)
\bigl(y_{1} + y_{2} - u_{\beta} - u_{{\cal P}_{\beta}} - v_{\beta} - v_{\tilde{{\cal P}}_{\beta}}\bigr)
+ z_{3} \bigl(y_{1} - u_{\beta} - u_{{\cal P}_{\beta}} \bigr)
\Bigr\}
\Biggr]
\times
\\
\nonumber
\\
\nonumber
&\times&
\prod_{\gamma=1}^{M-M_{1}}
\Biggl[
\int\int_{-\infty}^{+\infty} \frac{dqdy_{1}}{2\pi}
\Ai\bigl(y_{1} + q^{2} - f_{1}\bigr)
\exp\bigl\{i q \bigl(u_{M_{1} + \gamma} - u_{{\cal P}_{M_{1} + \gamma}}\bigr)\bigr\}
\times
\\
\nonumber
\\
&\times&
\int_{{\cal C}'} \frac{dz_{3}}{z_{3}}
\exp\bigl\{ z_{3} \bigl( y_{1} - u_{M_{1} + \gamma} - u_{{\cal P}_{M_{1}+\gamma} } \bigr)
\bigr\}
\Biggr]
\label{75}
\end{eqnarray}
Substituting here the explicit expression for the function $J_{*}\bigl(q- p, z_{1}, z_{2}, z_{3}\bigr)$,
eq.(\ref{66}), shifting the integrations over
$y_{1} \; \to \; y_{1} + u_{\alpha} + u_{{\cal P}_{\alpha}}$
and over
$y_{2} \; \to \; y_{2} + v_{\beta} + v_{\tilde{{\cal P}}_{\beta}}$,
and introducing summation over $M_{2} = M - M_{1}$ instead of the summation over $M$,
one can eventually represent the probability distribution function $W(f_{1},f_{2},\Delta)$
in the following sufficiently compact form:
\begin{eqnarray}
 \label{76}
W(f_{1},f_{2},\Delta) &=&
\sum_{M_{1}=0}^{\infty} \frac{(-1)^{M_{1}}}{(M_{1}!)^{2}}
\prod_{\beta=1}^{M_{1}}
\Biggl[
\int_{0}^{\infty} dv_{\beta} du_{\beta}
\Biggr]
\sum_{M_{2}=0}^{\infty} \frac{(-1)^{M_{2}}}{M_{2}!}
\;
\prod_{\gamma=1}^{M_{2}}
\Biggl[
\int_{0}^{\infty} du_{M_{1}+\gamma}
\Biggr]
\times
\\
\nonumber
\\
\nonumber
&\times&
\sum_{{\cal P}\in S_{M_{1}+M_{2}}} (-1)^{\bigl[{\cal P}\bigr]}
\sum_{\tilde{{\cal P}}\in S_{M_{1}}} (-1)^{\bigl[\tilde{{\cal P}}\bigr]}
\prod_{\beta=1}^{M_{1}}
\Bigl[
G\bigl(u_{\beta}, u_{{\cal P}_{\beta}}; \; v_{\beta}, v_{\tilde{{\cal P}}_{\beta}}\bigr)
\Bigr]
\times
\prod_{\gamma=1}^{M_{2}}
\Bigl[
A\bigl(u_{M_{1} + \gamma}, u_{{\cal P}_{M_{1} + \gamma}} \bigr)
\Bigr]
\end{eqnarray}
where
\begin{eqnarray}
 \nonumber
G\bigl(u, u'; \; v, v'\bigr) &=&
\int\int_{-\infty}^{+\infty} \frac{dqdp}{(2\pi)^{2}}
\int\int_{-\infty}^{+\infty} \frac{dy_{1} dy_{2}}{\Delta^{1/3}} \;
\Ai\bigl(y_{1} + q^{2} - f_{1} +u + u'\bigr) \; \exp\{iq(u-u')\}
\times
\\
\nonumber
\\
&\times&
\Ai\Bigl(\frac{y_{2} + \Delta \, p^{2} - (1+\Delta)^{1/3} f_{2} + f_{1} +v + v'}{\Delta^{1/3}}\Bigr)
\exp\{-ip(v-v')\} \;
{\cal S}\bigl(q- p, y_{1}, y_{2}\bigr)
\label{77}
\end{eqnarray}
with
\begin{eqnarray}
 \nonumber
 {\cal S}\bigl(q- p, y_{1}, y_{2}\bigr) &=&
\int_{{\cal C}} \frac{dz_{3}}{z_{3}}
\int_{{\cal C}} dz_{1}
\int_{{\cal C}} dz_{2}
\Bigl(\frac{1}{z_{1} z_{2}} - \delta(z_{1}) \delta(z_{2}) \Bigr) \;
\exp
\bigl\{
(z_{1} + z_{2})(y_{1} + y_{2}) \, + \, z_{3} y_{1}
\bigr\}
\times
\\
\nonumber
\\
&\times&
\Biggl(
1 + \frac{z_{1}}{z_{2} + \frac{1}{2} z_{3} + \frac{i}{2} (q-p)^{(-)}}
\Biggr)
\,
\Biggl(
1 + \frac{z_{2}}{z_{1} + \frac{1}{2} z_{3} - \frac{i}{2} (q-p)^{(+)}}
\Biggr)
\label{78}
\end{eqnarray}
and
\begin{eqnarray}
 \nonumber
A(u, \; u') &=&
\int_{-\infty}^{+\infty} dy \int_{-\infty}^{+\infty} \frac{dqd}{2\pi}
\Ai\bigl(y + q^{2} - f_{1} + u + u'\bigr)
\exp\bigl\{i q (u - u')\bigr\} \;
\int_{{\cal C}'} \frac{dz_{3}}{z_{3}}
\exp\{ z_{3} y \}
\\
\nonumber
\\
\nonumber
&=&
\int_{0}^{\infty} dy \int_{-\infty}^{+\infty} \frac{dq}{2\pi}
\Ai\bigl(y + q^{2} - f_{1} + u + u'\bigr)
\exp\bigl\{i q (u - u')\bigr\}
\\
\nonumber
\\
&=&
2^{1/3} K\Bigl[2^{1/3} u - f_{1}/2^{2/3}; \; 2^{1/3} u' - f_{1}/2^{2/3} \Bigr]
\label{79}
\end{eqnarray}
where $K_{s}(\omega, \omega')$ is the usual Airy kernel, eq.(\ref{14}).

Simple integrations over $z_{1}$, $z_{2}$ and $z_{3}$ in eq.(\ref{78}) yields:
\begin{equation}
{\cal S}\bigl(q- p, y_{1}, y_{2}\bigr) =
4\pi \, \delta(q-p) \, \delta(y_{1} + y_{2})  \; - \;
\theta(y_{1}) \, \theta(y_{2}) \; - \;
4 \delta(y_{1} + y_{2}) \, \theta(y_{1}) \,
\Bigl[\pi \delta(q-p) + \frac{\sin\bigl[(q-p) y_{1}\bigr]}{(q-p)}\Bigr]
\label{80}
\end{equation}

\vspace{10mm}

Eqs.(\ref{76}), (\ref{77}), (\ref{79}) and (\ref{80}) constitute the final result of the present research.
Note that although the obtained expression for the distribution function
$W(f_{1}, f_{2}, \Delta)$, eq.(\ref{76}), exhibits a "determinant-like" structure
it is {\it not} the Fredholm determinant. Here we are facing an object of
some other nature whose analytic properties are still to be investigated.


\subsection{The limit cases}

Substituting eq.(\ref{80}) into eq.(\ref{77}) and using the Airy function relations:
\begin{eqnarray}
\label{93}
&&\int_{-\infty}^{+\infty} \, dy \,
\Ai\bigl(y + u\bigr) \, \Ai\bigl(- \Delta^{-1/3} y + v \bigr) =
\frac{\Delta^{1/3}}{(1+\Delta)^{1/3}} \;
\Ai\Bigl[\frac{u + \Delta^{1/3} \, v}{(1+\Delta)^{1/3}} \Bigr]
\\
\nonumber
\\
\label{94}
&&\int_{-\infty}^{+\infty} \, \frac{dq}{2\pi}  \,
\Ai\bigl(a q^{2} + b\bigr) \;
\exp\bigl\{ iq c\bigr\}  =
2^{-1/3} \, a^{-1/2} \;
\Ai\bigl[2^{-2/3}(b + a^{-1/2} c)\bigr] \, \Ai\bigl[2^{-2/3}(b - a^{-1/2} c)\bigr]
\end{eqnarray}
the kernel $G(u, u' ; \; v, v')$, eq.(\ref{77}), can be represented in terms of three contributions:
\begin{equation}
 \label{96}
G\bigl( u, u' ; \; v, v'\bigr) \; = \;
\sum_{i=1}^{3} G_{i}\bigl( u, u' ; \; v, v'\bigr)
\end{equation}
where
\begin{eqnarray}
 \label{97}
G_{1}\bigl( u, u' ; \; v, v'\bigr) &=&
\frac{2^{2/3}}{\bigl( 1+ \Delta\bigr)^{2/3}}
\Ai\Biggl[\frac{2^{1/3}}{\bigl(1 + \Delta\bigr)^{1/3}} (u + v') - f_{2}/2^{2/3}\Biggr] \,
\Ai\Biggl[\frac{2^{1/3}}{\bigl(1 + \Delta\bigr)^{1/3}} (u' + v) - f_{2}/2^{2/3}\Biggr] \,
\\
\nonumber
\\
\nonumber
\\
\nonumber
G_{2}\bigl( u, u' ; \; v, v'\bigr)  &=&
 -\frac{2^{2/3}}{\Delta^{1/3}}
K\Bigl[2^{1/3} u - f_{1}/2^{2/3}; \; 2^{1/3} u' - f_{1}/2^{2/3} \Bigr]
\times
\\
\nonumber
\\
&\times&
K\Biggl[\frac{2^{1/3} v -  \bigl[\bigl(1+\Delta\bigr)^{1/3}f_{2} - f_{1}\bigr]/2^{2/3}}{\Delta^{1/3}}; \;
        \frac{2^{1/3} v' - \bigl[\bigl(1+\Delta\bigr)^{1/3}f_{2} - f_{1}\bigr]/2^{2/3}}{\Delta^{1/3}} \Biggr]
\label{98}
\end{eqnarray}
\begin{eqnarray}
\nonumber
G_{3}\bigl( u, u' ; \; v, v'\bigr)  &=&
 -\frac{4}{\Delta^{1/3}}
\int\int_{-\infty}^{+\infty} \frac{dqdp}{(2\pi)^{2}}
\int_{0}^{+\infty} dy \;
\Ai\Bigl(\frac{-y + \Delta \, p^{2} - (1+\Delta)^{1/3} f_{2} + f_{1} +v + v'}{\Delta^{1/3}}\Bigr)
\times
\\
\nonumber
\\
&\times&
\Ai\bigl(y + q^{2} - f_{1} + u + u'\bigr)
\Bigl[\pi \delta(q-p) + \frac{\sin\bigl[(q-p) y_{1}\bigr]}{(q-p)}\Bigr]
\exp\{iq(u-u') - ip(v-v')\}
\label{99}
\end{eqnarray}

Having explicit expressions for the integral kernels, eqs.(\ref{79}) and (\ref{97})-(\ref{99}),
one can study the properties of the distribution function $W(f_{1}, f_{2}, \Delta)$, eq.(\ref{76})
in the three limit cases:

\vspace{3mm}

{\bf (1) The limit }${\boldsymbol f_{2} \to -\infty}$.

In this case all three contributions (\ref{97})-(\ref{99}) turn to zero, so that
according to eq.(\ref{76}),
\begin{eqnarray}
\nonumber
\lim_{f_{2}\to -\infty} \; W\bigl(f_{1}, f_{2}, \Delta\bigr) &=&
\sum_{M_{2}=0}^{\infty} \frac{(-1)^{M_{2}}}{M_{2}!}
\;
\prod_{\gamma=1}^{M_{2}}
\Biggl[
\int_{0}^{\infty} du_{\gamma}
\Biggr]
\times
\\
\nonumber
\\
\nonumber
&\times&
\sum_{{\cal P}\in S_{M_{2}}} (-1)^{\bigl[{\cal P}\bigr]}
\prod_{\gamma=1}^{M_{2}}
\Bigl[
A\bigl(u_{\gamma}, u_{{\cal P}_{\gamma}} \bigr)
\Bigr]
\\
\nonumber
\\
&=&
\sum_{M=0}^{\infty} \frac{(-1)^{M}}{M!}
\;
\prod_{\gamma=1}^{M}
\Biggl[
\int_{0}^{\infty} du_{\gamma}
\Biggr]
\det\Bigl[A\bigl(u_{\gamma}, u_{\gamma'} \bigr)\Bigr]_{\gamma, \gamma' = 1, ..., M}
\label{100}
\end{eqnarray}
Substituting here eq.(\ref{79}) we get
\begin{equation}
\label{101}
\lim_{f_{2}\to -\infty} \; W\bigl(f_{1}, f_{2}, \Delta\bigr) \; = \;
F_{2}\bigl(-f_{1}/2^{2/3}\bigr)
\end{equation}
which is the GUE Tracy-Widom distribution for $f_{1}$, as it should be.

\vspace{3mm}

{\bf (2) The limit} ${\boldsymbol  f_{1} \to -\infty}$.

In this limit the kernels $G_{2}(u, u'; \; v, v')$, eq.(\ref{98}), $G_{3}(u, u'; \; v, v')$, eq.(\ref{99}),
and $A(v, v')$, eq.(\ref{79}) turn to zero. Substituting the kernel $G_{1}(u, u'; \; v, v')$, eq.(\ref{97}),
into eq.(\ref{76}), we get
\begin{eqnarray}
\nonumber
\lim_{f_{1}\to -\infty} \; W\bigl(f_{1}, f_{2}, \Delta\bigr) &=&
\sum_{M_{1}=0}^{\infty} \frac{(-1)^{M_{1}}}{(M_{1}!)^{2}}
\prod_{\beta=1}^{M_{1}}
\Biggl[
\int_{0}^{\infty} dv_{\beta} du_{\beta}
\Biggr]
\sum_{{\cal P}\in S_{M_{1}}} (-1)^{\bigl[{\cal P}\bigr]}
\sum_{\tilde{{\cal P}}\in S_{M_{1}}} (-1)^{\bigl[\tilde{{\cal P}}\bigr]}
\times
\\
\nonumber
\\
\nonumber
&\times&
\prod_{\beta=1}^{M_{1}}
\Biggl\{
\frac{2^{2/3}}{\bigl( 1+ \Delta\bigr)^{2/3}}
\Ai\Biggl[
\frac{2^{1/3}}{\bigl(1 + \Delta\bigr)^{1/3}} (u_{\beta} + v_{\tilde{{\cal P}}_{\beta}}) - f_{2}/2^{2/3}
\Biggr] \,
\Ai\Biggl[
\frac{2^{1/3}}{\bigl(1 + \Delta\bigr)^{1/3}} (u_{{\cal P}_{\beta}} + v_{\beta}) - f_{2}/2^{2/3}
\Biggr]
\Biggr\}
\label{102}
\end{eqnarray}
redefining, $u_{\beta} \to 2^{-1/3} \bigl(1 + \Delta\bigr)^{1/3} \; u_{\beta}$ and
$v_{\beta} \to 2^{-1/3} \bigl(1 + \Delta\bigr)^{1/3} \; v_{\beta}$, and taking into account that
\begin{equation}
 \label{103}
\prod_{\beta=1}^{M}
\Ai\Bigl[u_{{\cal P}_{\beta}} + v_{\beta} - f_{2}/2^{2/3}\Biggr] \; = \;
\prod_{\beta=1}^{M}
\Ai\Bigl[u_{\beta} + v_{{\cal P}^{-1}_{\beta}} - f_{2}/2^{2/3}\Biggr]
\end{equation}
we obtain
\begin{eqnarray}
\nonumber
\lim_{f_{1}\to -\infty} \; W\bigl(f_{1}, f_{2}, \Delta\bigr) &=&
\sum_{M=0}^{\infty} \frac{(-1)^{M}}{(M!)^{2}}
\prod_{\beta=1}^{M}
\Biggl[
\int_{0}^{\infty} dv_{\beta}
\Biggr]
\sum_{{\cal P}\in S_{M}} (-1)^{\bigl[{\cal P}\bigr]}
\sum_{\tilde{{\cal P}}\in S_{M}} (-1)^{\bigl[\tilde{{\cal P}}\bigr]}
\times
\\
\nonumber
\\
&\times&
\prod_{\beta=1}^{M}
\Biggl\{
\int_{0}^{\infty} du_{\beta} \;
\Ai\Bigl[u_{\beta} + v_{\tilde{{\cal P}}_{\beta}} - f_{2}/2^{2/3}\Bigr] \,
\Ai\Bigl[u_{\beta} + v_{{\cal P}^{-1}_{\beta}} - f_{2}/2^{2/3}\Bigr]\Biggr\}
\label{104}
\end{eqnarray}
integrating over $u_{\beta}$ and taking into account that
\begin{equation}
 \label{105}
\prod_{\beta=1}^{M}
\Bigl[ K\bigl(v_{{\cal P}^{-1}_{\beta}} - f_{2}/2^{2/3}; \;
v_{\tilde{{\cal P}}_{\beta}} - f_{2}/2^{2/3} \bigr)
\Bigr]
\; = \;
\prod_{\beta=1}^{M}
\Bigl[ K\bigl(v_{\beta} - f_{2}/2^{2/3} ; \;
       v_{({\cal P} + \tilde{{\cal P}})_{\beta}} - f_{2}/2^{2/3}\bigr)
\Bigr]
\end{equation}
we can redefine ${\cal P} + \tilde{{\cal P}} \; =  \; {\cal P}'$. In this way the
above expression becomes independent of the permutations $\tilde{{\cal P}}$ which
provide the factor $M!$ in eq.(\ref{104}). Thus,
\begin{eqnarray}
\nonumber
\lim_{f_{1}\to -\infty} \; W\bigl(f_{1}, f_{2}, \Delta\bigr) &=&
\sum_{M=0}^{\infty} \frac{(-1)^{M}}{M!}
\prod_{\beta=1}^{M}
\Biggl[
\int_{0}^{\infty} dv_{\beta}
\Biggr]
\sum_{{\cal P}'\in S_{M}} (-1)^{\bigl[{\cal P}'\bigr]}
\;
\prod_{\beta=1}^{M}
\Bigl[
K\bigl(v_{\beta} - f_{2}/2^{2/3} ; \;
       v_{{\cal P}'_{\beta}} - f_{2}/2^{2/3}\bigr)
\Bigr]
\\
\nonumber
\\
&=&
F_{2}\bigl(-f_{2}/2^{2/3}\bigr)
\label{106}
\end{eqnarray}
which is the GUE Tracy-Widom distribution for $f_{2}$, as it should be.

\vspace{3mm}

{\bf (3) The limit }${\boldsymbol \Delta \to \infty}$.

In this case the kernels $G_{2}(u, u'; \; v, v')$, eq.(\ref{98}), $G_{3}(u, u'; \; v, v')$, eq.(\ref{99}),
turn to zero. Substituting $A(v, v')$, eq.(\ref{79}), and $G_{1}(u, u'; \; v, v')$, eq.(\ref{97}),
into eq.(\ref{76}),  we get
\begin{eqnarray}
\nonumber
\lim_{\Delta\to \infty} \; W\bigl(f_{1}, f_{2}, \Delta\bigr) &=&
\lim_{\Delta\to \infty}
\sum_{M_{1}=0}^{\infty} \frac{(-1)^{M_{1}}}{(M_{1}!)^{2}}
\prod_{\beta=1}^{M_{1}}
\Biggl[
\int_{0}^{\infty} dv_{\beta} du_{\beta}
\Biggr]
\sum_{M_{2}=0}^{\infty} \frac{(-1)^{M_{2}}}{M_{2}!}
\;
\prod_{\gamma=1}^{M_{2}}
\Biggl[
\int_{0}^{\infty} du_{M_{1}+\gamma}
\Biggr]
\times
\\
\nonumber
\\
\nonumber
&\times&
\sum_{{\cal P}\in S_{M_{1}+M_{2}}} (-1)^{\bigl[{\cal P}\bigr]}
\sum_{\tilde{{\cal P}}\in S_{M_{1}}} (-1)^{\bigl[\tilde{{\cal P}}\bigr]}
\times
\\
\nonumber
\\
\nonumber
&\times&
\prod_{\beta=1}^{M_{1}}
\Biggl\{
\frac{2^{2/3}}{\Delta^{2/3}}
\Ai\Biggl[
\frac{2^{1/3}}{\Delta^{1/3}} (u_{\beta} + v_{\tilde{{\cal P}}_{\beta}}) - f_{2}/2^{2/3}
\Biggr] \,
\Ai\Biggl[
\frac{2^{1/3}}{\Delta^{1/3}} (u_{{\cal P}_{\beta}} + v_{\beta}) - f_{2}/2^{2/3}
\Biggr]\Biggr\}
\times
\\
\nonumber
\\
&\times&
\prod_{\gamma=1}^{M_{2}}
\Biggl\{
2^{1/3} K\Bigl[2^{1/3} u_{M_{1}+\gamma} - f_{1}/2^{2/3}; \; 2^{1/3} u_{{\cal P}_{M_{1} + \gamma}} - f_{1}/2^{2/3} \Bigr]
\Biggr\}
\label{107}
\end{eqnarray}
Performing the transformations similar to the ones described above
in eq.(\ref{102})-(\ref{106}), we find
\begin{eqnarray}
\nonumber
\lim_{\Delta\to \infty} \; W\bigl(f_{1}, f_{2}, \Delta\bigr) &=&
\lim_{\Delta\to \infty}
\sum_{M_{1}=0}^{\infty} \frac{(-1)^{M_{1}}}{M_{1}!}
\prod_{\beta=1}^{M_{1}}
\Biggl[
\int_{0}^{\infty}du_{\beta}
\Biggr]
\sum_{M_{2}=0}^{\infty} \frac{(-1)^{M_{2}}}{M_{2}!}
\;
\prod_{\gamma=1}^{M_{2}}
\Biggl[
\int_{0}^{\infty} du_{M_{1}+\gamma}
\Biggr]
\sum_{{\cal P}\in S_{M_{1}+M_{2}}} (-1)^{\bigl[{\cal P}\bigr]}
\times
\\
\nonumber
\\
\nonumber
&\times&
\prod_{\beta=1}^{M_{1}}
\Biggl\{
\frac{2^{1/3}}{\Delta^{1/3}}
K\Biggl[
\frac{2^{1/3}}{\Delta^{1/3}} u_{\beta} - f_{2}/2^{2/3} ; \;
\frac{2^{1/3}}{\Delta^{1/3}} u_{{\cal P}_{\beta}}  - f_{2}/2^{2/3}
\Biggr]\Biggr\}
\times
\\
\nonumber
\\
&\times&
\prod_{\gamma=1}^{M_{2}}
\Biggl\{
2^{1/3} K\Bigl[2^{1/3} u_{M_{1}+\gamma} - f_{1}/2^{2/3}; \; 2^{1/3} u_{{\cal P}_{M_{1} + \gamma}} - f_{1}/2^{2/3} \Bigr]
\Biggr\}
\label{108}
\end{eqnarray}
Redefining $u_{\beta} \to 2^{-1/3} \Delta^{1/3} u_{\beta}$ for $\beta = 1, ..., M_{1}$
and $u_{\gamma} \to 2^{-1/3} u_{\gamma}$ for $\gamma = M_{1} + 1, ..., M_{1} + M_{2}$, we find that
in the limit $\Delta \to \infty$ the summation over the permutations ${\cal P} \in S_{M_{1}+M_{2}}$
in eq.(\ref{108}) splits into two independent summations over ${\cal P}^{(1)} \in S_{M_{1}}$
and ${\cal P}^{(2)} \in S_{M_{2}}$ (one can easily see that after the above rescaling, all the factors \\
$K\bigl(2^{1/3} u_{M_{1}+\gamma} - f_{1}/2^{2/3}; \; 2^{1/3} u_{{\cal P}_{M_{1} + \gamma}} - f_{1}/2^{2/3} \bigr)$
in which $u_{{\cal P}_{M_{1} + \gamma}}$ turns out to be one of the  parameters
in the set $ \{\Delta^{1/3} u_{1}, \Delta^{1/3} u_{2}, ..., \Delta^{1/3} u_{M_{1}}\}$
turn to zero in the limit $\Delta \to \infty$).
Thus,
\begin{eqnarray}
\nonumber
\lim_{\Delta\to \infty} \; W\bigl(f_{1}, f_{2}, \Delta\bigr) &=&
\sum_{M_{1}=0}^{\infty} \frac{(-1)^{M_{1}}}{M_{1}!}
\prod_{\beta=1}^{M_{1}}
\Biggl[
\int_{0}^{\infty}du_{\beta}
\Biggr]
\sum_{M_{2}=0}^{\infty} \frac{(-1)^{M_{2}}}{M_{2}!}
\;
\prod_{\gamma=1}^{M_{2}}
\Biggl[
\int_{0}^{\infty} du_{M_{1}+\gamma}
\Biggr]
\times
\\
\nonumber
\\
\nonumber
&\times&
\sum_{{\cal P}^{(1)}\in S_{M_{1}}} (-1)^{[{\cal P}^{(1)}]}
\sum_{{\cal P}^{(2)}\in S_{M_{2}}} (-1)^{[{\cal P}^{(2)}]}
\times
\\
\nonumber
\\
\nonumber
&\times&
\prod_{\beta=1}^{M_{1}}
\Bigl[
K\bigl(
u_{\beta} - f_{2}/2^{2/3} ; \;
u_{{\cal P}^{(1)}_{\beta}}  - f_{2}/2^{2/3}
\bigr)
\Bigr]
\prod_{\gamma=1}^{M_{2}}
\Bigl[
K\bigl(u_{M_{1}+\gamma} - f_{1}/2^{2/3}; \; u_{M_{1} + {\cal P}^{(2)}_{\gamma}} - f_{1}/2^{2/3} \bigr)
\Bigr]
\\
\nonumber
\\
&=&
F_{2}\bigl[-f_{1}/2^{2/3}\bigr] \, F_{2}\bigl[-f_{2}/2^{2/3}\bigr]
\label{109}
\end{eqnarray}
Which is the product of two independent GUE Tracy-Widom distributions for $f_{1}$ and $f_{2}$
as it should be.

\vspace{3mm}

The limit $\Delta \to 0$ is much more tricky. First of all, technically it is not so easy
to study, and second it is not quite clear what kind of behavior  for the probability
distribution function should be expected in this case, as the result for the function
$W\bigl(f_{1}, f_{2}, \Delta\bigr)$
has been derived in the limit $(t_{2} - t_{1}) = \Delta \, t \to \infty$. In this situation
the physical meaning of the limit $\Delta \to 0$ is not evident.

\section{Conclusions}

In this paper the analytic expression for the two time free energy distribution function
in (1+1) directed polymers with the zero boundary conditions,
$W(f_{1}, f_{2}, \Delta) = \lim_{t\to\infty}
\mbox{Prob}\bigl[f_{t}(0) > f_{1}; \; f_{t+\Delta t}(0) > f_{2} \bigr]$
has been derived. It should be stressed the obtained result,
eqs.(\ref{76}), (\ref{77}), (\ref{79}) and (\ref{80}), should be considered
as the preliminary one. At present stage it is expressed in terms rather
complicated "determinant-like" object, eq.(\ref{76}), whose analytic properties are still to be studied,
although in the three  limit cases the derived expression reduces to
predictable results, namely: \\
$\lim_{f_{1}\to -\infty} \; W\bigl(f_{1}, f_{2}, \Delta\bigr) =
F_{2}\bigl(-f_{2}/2^{2/3}\bigr)$;
$\; \lim_{f_{2}\to -\infty} \; W\bigl(f_{1}, f_{2}, \Delta\bigr) =
F_{2}\bigl(-f_{1}/2^{2/3}\bigr)$;
$\; \lim_{\Delta \to -\infty} \; W\bigl(f_{1}, f_{2}, \Delta\bigr)  =
F_{2}\bigl(-f_{1}/2^{2/3}\bigr) F_{2}\bigl(-f_{2}/2^{2/3}\bigr)$.

Note finally that the obtained result can be easily generalized for the case when
the directed polymer at time $t$ comes not to zero but to a given point $x_{1}$,
while at time $t + \Delta t$ it comes to another given point $x_{2}$. After the proper rescaling
$x_{1,2} \to \bigl(\beta u t^{2} \bigr)^{1/3} \, x_{1,2}$ the expression for the
corresponding distribution function $ \tilde{W}\bigl(f_{1}, x_{1}, f_{2}, x_{2}, \Delta\bigr)$
is obtained from $W\bigl(f_{1}, f_{2}, \Delta\bigr)$ by the trivial shift:
\begin{equation}
\label{110}
\tilde{W}\bigl(f_{1}, x_{1}, f_{2}, x_{2}, \Delta\bigr) \; = \;
W\bigl(f_{1} + x_{1}^{2}/2, \, f_{2} + x_{2}^{2}/2, \, \Delta\bigr)
\end{equation}

\acknowledgments

An essential part of this work was carried out during the Symposium on
the KPZ-equation (February 24 - March 2, 2013)
at St.John, Virgin Islands, funded by the Simons Foundation.

I am grateful to Alexei Borodin, Kostya Khanin, Herbert Spohn, Jeremy Quastel,
Senya Shlosman, Patrick Ferrari, Pierre Le Doussal, Pasquale Calabrese and
Ivan Corwin for numerous illuminating discussions which were crucial for
the progress in these somewhat complicated calculations.

This work was supported in part by the grant IRSES DCPA PhysBio-269139.



\end{document}